\documentclass[12pt]{article}
\usepackage{graphicx}
\usepackage{amssymb}
\usepackage{verbatim}
\advance\textheight by \topskip
\begin{document}
\tolerance=100000
\thispagestyle{empty}
\setcounter{page}{0}

\def\cO#1{{\cal{O}}\left(#1\right)}
\newcommand{\be}{\begin{equation}}
\newcommand{\ee}{\end{equation}}
\newcommand{\br}{\begin{eqnarray}}
\newcommand{\er}{\end{eqnarray}}
\newcommand{\ba}{\begin{array}}
\newcommand{\ea}{\end{array}}
\newcommand{\bi}{\begin{itemize}}
\newcommand{\ei}{\end{itemize}}
\newcommand{\bn}{\begin{enumerate}}
\newcommand{\en}{\end{enumerate}}
\newcommand{\bc}{\begin{center}}
\newcommand{\ec}{\end{center}}
\newcommand{\ul}{\underline}
\newcommand{\ol}{\overline}
\newcommand{\ra}{\rightarrow}
\newcommand{\sm}{SM~}
\newcommand{\as}{\alpha_s}
\newcommand{\aem}{\alpha_{em}}
\newcommand{\ta}{\theta_1}
\newcommand{\tb}{\theta_2}
\newcommand{\ga}{\gamma_a}
\newcommand{\gb}{\gamma_b}
\newcommand{\bea}{\beta_1}
\newcommand{\beb}{\beta_2}
\newcommand{\cph}{c_\phi}

\catcode`\@=11

\def\@citex[#1]#2{\if@filesw\immediate\write\@auxout{\string\citation{#2}}\fi
  \def\@citea{}\@cite{\@for\@citeb:=#2\do
    {\@citea\def\@citea{,\penalty\@m}\@ifundefined
       {b@\@citeb}{{\bf ?}\@warning
       {Citation `\@citeb' on page \thepage \space undefined}}%
\hbox{\csname b@\@citeb\endcsname}}}{#1}}

\def\citer{\@ifnextchar [{\@tempswatrue\@citexr}{\@tempswafalse\@citexr[]}}
%

\def\@citexr[#1]#2{\if@filesw\immediate\write\@auxout{\string\citation{#2}}\fi
  \def\@citea{}\@cite{\@for\@citeb:=#2\do
    {\@citea\def\@citea{--\penalty\@m}\@ifundefined
       {b@\@citeb}{{\bf ?}\@warning
       {Citation `\@citeb' on page \thepage \space undefined}}%
\hbox{\csname b@\@citeb\endcsname}}}{#1}}
\catcode`\@=12
\def\Ecm{\ifmmode{E_{\mathrm{cm}}}\else{$E_{\mathrm{cm}}$}\fi}
\def\lsim{\buildrel{\scriptscriptstyle <}\over{\scriptscriptstyle\sim}}
\def\gsim{\buildrel{\scriptscriptstyle >}\over{\scriptscriptstyle\sim}}
\def \lum{{\cal L}}

\def\lapp{\mathrel{\rlap{\raise.5ex\hbox{$<$}}
                    {\lower.5ex\hbox{$\sim$}}}}
\def\gapp{\mathrel{\rlap{\raise.5ex\hbox{$>$}}
                    {\lower.5ex\hbox{$\sim$}}}}
\newcommand{\decay}[2]{
\begin{picture}(25,20)(-3,3)
\put(0,-20){\line(0,1){10}}
\put(0,-20){\vector(1,0){15}}
\put(0,0){\makebox(0,0)[lb]{\ensuremath{#1}}}
\put(25,-20){\makebox(0,0)[lc]{\ensuremath{#2}}}
\end{picture}}
\vspace*{\fill}
\begin{flushright}
{IISc-CHEP/08/07}\\
{CERN-PH-TH/2007-116}\\
{LAPTH-1195/07}\\
\end{flushright}
\vspace{1cm}
\begin{center}
{\Large \bf
Aspects of CP violation in the \boldmath{$H ZZ$} coupling at the LHC}
 \\[0.25cm]
\end{center}
\begin{center}
{\large  Rohini M. Godbole${^{a}}$, David J. Miller${^{b}}$
and M.~Margarete~M\"uhlleitner$^{c,d}$ }\\[0.35 cm]
$^a$Centre for High Energy Physics, Indian Institute of Science, Bangalore, 560 012, India.\\[0.20cm]
$^b$ Dept. of Physics and Astronomy, University of Glasgow, Glasgow G12 8QQ, U.K.\\[0.20cm]
$^c$ Theory Division, Physics Department, CERN, CH-1211 Geneva 23,
Switzerland.\\[0.20cm]
$^d$ Laboratoire d'Annecy-Le-Vieux de Physique Th\'eorique, LAPTH, France.
\end{center}

\vspace{.8cm}

\begin{abstract}
{\noindent\normalsize We examine the CP-conserving (CPC) and
CP-violating (CPV) effects of a general $H ZZ$ coupling
through a study of the process $H \ra Z Z^{(*)}
\ra \ell^+ \ell^- \ell^{'+} \ell^{'-}$ at the LHC.  We construct
asymmetries that directly probe these couplings.  Further, we present
complete analytical formulae for the angular distributions of the
decay leptons and for some of the asymmetries. Using these we have been able 
to identify new observables which can provide enhanced sensitivity to the 
CPV $H ZZ$ coupling. We also explore probing CP violation through shapes of 
distributions in different kinematic variables, which can be used for Higgs
bosons with $m_H < 2\, m_Z$.  }
\end{abstract}
\vskip 1.0cm
\noindent
\vspace*{\fill}
\newpage

\section{Introduction}
\label{intro}

The Standard Model (SM) has had unprecedented success in passing
precision tests at the SLC, LEP, HERA and the Tevatron. However, the
verification of the Higgs mechanism, which allows the generation of
particle masses for fermions and electroweak (EW) gauge bosons without
violating the gauge principle, is still lacking. The search for the
Higgs boson and the study of its properties will be among the major
tasks of the Large Hadron Collider (LHC), which will soon start
operation, and of the International Linear Collider (ILC), which is
under planning and consideration~\cite{myreview}.

However, the instability of the Higgs boson mass to radiative
corrections and the resulting fine tuning problem point towards the
existence of physics beyond the SM (BSM) at the TeV scale. This BSM
physics usually implies more Higgs bosons and may have implications for 
the properties of the Higgs boson(s). Hence, the determination of the Higgs 
boson quantum numbers and properties will be crucial to establish it as 
{\it the} SM Higgs boson~\cite{abdel-tome} or to probe 
any new BSM physics.

Furthermore, there is no real theoretical understanding of the
relative magnitudes and phases of the different fermion mass
parameters in the SM, even though we have an extremely successful
description of {\it all} observed CP-violation (CPV) in terms of the
Cabbibo-Kobayashi-Masakawa (CKM) matrix. Indeed, the CPV of the SM,
observed only in the $K_0$--$\bar K_0$ and $B_0$--$\bar B_0$ systems
to date, appears insufficient to explain the Baryon Asymmetry of the
Universe (BAU)~\cite{baryoreview}, and an additional source of CPV
beyond that of the \sm may be needed for a {\it quantitative}
explanation. An extended Higgs sector together with CPV supersymmetry
(SUSY) is one possible BSM option that may explain this
BAU~\cite{nmssmbaryo}. Thus it is clear that the knowledge of the
properties of the Higgs sector and any possible CPV therein is of
utmost importance in particle physics phenomenology at
present~\cite{CPNSHrep,mypramreview}.

The LHC will search for the SM Higgs boson in the entire mass range
expected theoretically and still allowed experimentally~\cite{tdratl,tdrcms},
whereas precision profiling of the Higgs boson is expected to be one
of the focal points at the ILC~\cite{ilc}. After discovery, the 
determination of the Higgs boson couplings, in particular
those with a pair of electroweak gauge bosons $(V=W/Z)$ and those with a
pair of heavy fermions $(f = t/\tau)$, will be essential. In this study
we focus on the $H ZZ$ coupling.

The ILC, in both the $e^+e^-$ and the $\gamma \gamma$ \cite{photoncoll} 
options, and the
LHC offer a wealth of possibilities for the exploration of the CP
quantum numbers of the Higgs boson $H$~\cite{Godbole:2004xe}. 
At an $e^+e^-$ collider, the $Z$ boson produced in the process $e^+ e^- \ra
Z H$ is at high energies longitudinally polarised when produced in 
association with a CP-even Higgs boson and transversely polarised in case
of a CP-odd Higgs boson. The angular
distribution of the $Z$ boson therefore carries a footprint of the
Higgs boson's CP properties~\citer{Barger:1993wt,CP-full}.
Furthermore, measurements of the threshold excitation curve can yield
useful information on the spin and the parity of the Higgs boson and
establish it to have spin 0 and be even under parity transformation,
hence $J^P = 0^{+}$, in a model-independent 
way~\cite{Miller:2001bi,Dova:2003py}.
Additionally, kinematic distributions of the final state particles in
the process $e^+ e^- \to f \bar f H$, produced via vector boson fusion
or Higgsstrahlung, where $f$ is a light fermion, with or without
initial beam polarisation, can be exploited to study the $H ZZ$
coupling, including CPV~\cite{Hagiwara:1993sw}, \citer{dist,Biswal:2005fh}.
Ref.~\cite{Hagiwara:2000tk} uses the optimal observable technique
whereas Refs.~\cite{Chang:1993jy,Han:2000mi,Biswal:2005fh} exploit the
kinematical distributions to construct asymmetries that are directly
proportional to different parts of a general CP-violating
coupling. Associated production with top quarks $e^+ e^- \to t \bar t
H$ may be used to extract CP information too~\cite{Gunion:1996vv,bhupal}.

Higgs decays may also be used effectively.  The angular distributions
of the Higgs decay products, either a pair of vector bosons or heavy
fermions that further decay, can be exploited to gain information on
the Higgs CP properties if it is a CP-eigenstate and the CP-mixing if
it is CP violating~\cite{Chang:1993jy}, \citer{Kramer:1993jn,Choi:2002jk}.
A detailed study of the Higgs spin and parity using the angular
distributions of the final-state fermions in $H \to ZZ \to$~{\it
leptons}, above and below the $ZZ$ threshold, was performed in
\cite{Choi:2002jk}.
The $H \to f \bar f$ pair $(f = t/\tau)$ has the advantage of being
equally sensitive to the CP-even and CP-odd part of the Higgs
boson~\cite{Bower:2002zx}.
For Higgs bosons produced in association with heavy fermions, or Higgs
decays to heavy fermions at an $e^+e^-$ collider, angular correlations
and/or the polarisations of the heavy fermions may also be
used~\cite{bhupal,Grzadkowski:1995rx,chinese}.

An ILC operating in the $\gamma\gamma$ mode offers
an attractive option not only for the CP-determination of the Higgs
boson, but also for the measurement of a small CP-mixing in a state
that is dominantly CP-even.  Using linear and circular polarisation of
the photons one can get a clear measure of the CP
mixing~\cite{Grzadkowski:1992sa}; further using a circular beam
polarization, the almost mass degenerate CP-odd and CP-even
Higgs bosons of the MSSM may be separated
\citer{Muhlleitner:2001kw,asner}.
Interference effects in the process $\gamma \gamma \rightarrow
H \rightarrow f \bar f \; (f = t/\tau)$
\citer{Anlauf:1995mu,Ellis:2004hw} can be
used to determine the $f \bar f H$ and $\gamma \gamma H$ couplings for
an $H$ with indefinite CP parity. 

Hence, the $e^+e^-$ collider and its possible operation as a $\gamma
\gamma$ collider offer some unique possibilities in the exploration of
the CP quantum numbers of the Higgs boson. However, the LHC is the
next collider to come into operation. So we want to seek answers to
these questions already at the LHC~\cite{Hohl:2001atl}.  Here, the
$t\bar t$ final state produced in the decay of an inclusively produced
Higgs boson can provide knowledge of the CP nature of the $t\bar t H$
coupling through spin-spin
correlations~\cite{Bernreuther:1997gs,Khater:2003wq} whereas $t\bar t
H$ production allows a determination of the CP-even and CP-odd part of
the $f \bar f$ couplings with the Higgs boson
separately~\cite{Gunion:1996xu,Field:2002gt}.  The use of $\tau$
polarisation in resonant $\tau^+ \tau^-$ production at the LHC has
also been recently investigated~\cite{Ellis:2004fs}.  The $H ZZ$
coupling can be explored at the LHC in the Higgs decay into a $Z$
boson pair which then decay each into a lepton pair, {\it i.e.}  $H
\to ZZ^{(*)} \to (\ell^+ \ell^-) (\ell^{'+}
\ell^{'-})$~\cite{Choi:2002jk},
\citer{Buszello:2002uu,Allanach:2006yt}; above threshold, angular
distributions have to be used while below threshold, the dependence on
the virtual $Z^*$ boson's invariant mass may be
exploited. Furthermore, this coupling (and the $HWW$ coupling) can be
studied in vector boson fusion~\citer{Plehn:2001nj,Buszello:2006hf},
and a similar idea may be employed in $H + 2$~jet
production~\cite{DelDuca:2001ad,Hankele:2006ja} in gluon fusion
(however, also see Ref.\cite{Odagiri:2002nd}).

Most of the suggested measurements should be able to verify a scalar
Higgs boson when the full luminosity of $300\,$fb$^{-1}$ is collected
at the LHC (or even before), provided the Higgs boson is a CP
eigenstate.  For example, using the threshold behaviour it may be possible 
to rule out a pure pseudoscalar state with $100\,$fb$^{-1}$ in the
SM~\cite{Choi:2002jk}.
However, a measurement of the CP mixing is
much more difficult, and a combination of several different
observables will be essential.

In this paper we investigate CP mixing in the Higgs sector using the
process, \mbox{$H \ra Z Z^{(*)} \ra (\ell^+ \ell^-) (\ell^{'+} \ell^{'-})$}.  
We extend the analysis of Ref.~\cite{Choi:2002jk} to a
Higgs boson of indefinite CP. Further, we extend the analysis of
Ref.~\cite{Allanach:2006yt}, where asymmetries were constructed using
angular distributions of the decay leptons, which directly probe the
CP mixing. 

The paper is organised as follows. In section~\ref{secttwo} we present
the complete analytical formulae for the angular distribution of the
decay leptons produced in the process $H \ra Z Z^{(*)} \ra 
(\ell^+ \ell^-) (\ell^{'+} \ell^{'-})$, parameterising the $H ZZ$ vertex in a
model-independent way, for a Higgs boson of indefinite CP. In
section~\ref{sectthree} we examine how this modified coupling changes
the total number of $H \to ZZ \to 4\,${\it lepton} events seen at the
LHC. In section~\ref{sectfour} we then construct different observables that 
can be used to probe the CP nature of the Higgs boson and present the numerical
results. In section~\ref{sectfive}, we propose an investigation of CP
mixing using kinematical distributions of the decay leptons, and in
section~\ref{sectconcl} we present our conclusions.\\

\section{Model independent analysis of \boldmath{$H \ra ZZ^{(*)}$}}
\label{secttwo}

For our study of possible CPV in the Higgs sector we will
examine the decay of a Higgs boson into two $Z$ bosons
with subsequent decay into two lepton pairs,
\br
H \to ZZ^{(*)} \to (f_1 \bar{f}_1) (f_2 \bar{f}_2) \;.
\label{hdecayzz}
\er
To perform a model-independent analysis we examine the most general
vertex including possible CPV for a spin-0 boson\footnote{In fact, in
order to be as general as possible one should allow for a general CP
violating coupling with a ``Higgs'' particle of arbitrary spin, as in
\cite{Choi:2002jk}. We keep this for future work.} coupling to two $Z$
bosons with four-momenta $q_1$ and $q_2$, respectively. This can be
written as
\br V_{HZZ}^{\mu \nu} \, =\,
\frac{ig m_Z}{\cos\theta_W} \left[ \,a\, g_{\mu\nu} 
+  b \,\frac{p_\mu p_\nu}{m_Z^2}  
+  c \,\epsilon_{\mu\nu\alpha\beta} \, \frac{ p^\alpha k^\beta}{m_Z^2} 
\, \right],
\label{param}
\er
where $p=q_1+q_2$ and $k=q_1-q_2$, $\theta_W$ denotes the weak-mixing angle
and $\epsilon_{\mu \nu\alpha\beta}$ is the totally antisymmetric tensor with
$\epsilon_{0123}=1$. As can be inferred from Eq.~(\ref{param}) the CP 
conserving tree-level Standard  Model coupling is recovered for $a=1$ and 
$b=c=0$.

The terms containing $a$ and $b$ are associated with the coupling of a 
CP-even Higgs boson to a pair of $Z$ bosons, while
that containing $c$ is associated with that of a CP-odd Higgs boson. In
general these parameters can be momentum-dependent form
factors that may be generated from loops containing new heavy
particles or equivalently from the integration over heavy degrees of
freedom giving rise to higher dimensional operators.
The form factors $b$ and $c$ may, in general, be complex. Since an
overall phase will not affect the observables studied here, we are
free to adopt the convention that $a$ is real. This convention
requires the assumption that the signal and background do not
interfere, and indeed in our approximation where the Higgs boson is
taken on-shell, this interference is exactly zero. Interference would
be only manifest if the Higgs boson were taken off-shell and since the
dominant signal contribution arises from on-shell Higgs bosons, we
expect this interference to be small and neglect it.

In principle, the vertex is valid at all orders in perturbation
theory.  Contributions to the $HZZ$ vertex from loop corrections will
not add any new tensor structures and will only alter the values of
$a$, $b$ and $c$. More generally, $a$, $b$ and $c$ are momentum
dependent form factors obtained from integrating out the new physics
at some large scale $\Lambda$. Since the momentum dependence will
involve ratios of typical momenta in the process to the large scale
$\Lambda$, we make the reasonable assumption that the scale dependence
can be neglected and keep only the constant part.

Non-vanishing values for either $\Im m(b)$ or $\Im m(c)$ destroy the
hermiticity of the effective theory. Such couplings can be envisaged
when going beyond the Born approximation, where they arise from
final state interactions, or, in other words out of absorptive parts
of the higher order diagrams, presumably mediated by new
physics. Further, $a$, $\Re e(b)$ and $\Im m(c)$ are even under
$\tilde{\rm T}$, while $\Im m(b)$ and $\Re e(c)$ are odd, where
$\tilde{\rm T}$ stands for the pseudo-time reversal transformation, which
reverses particle momenta and spins but does not interchange initial
and final states.  It is the ${\rm CP\tilde T}$ odd coefficients that
are related to the presence of absorptive parts in the
amplitude~\cite{Hagiwara:1986vm}. In most CPV extensions of the SM one
has $|a| \gg |b|, |c|$, so most of the observables used to study the
$HZZ$ vertex are dominated by the first term in the vertex Eq.~(\ref{param}); 
in order to probe the last, the CP-odd term, it is most advantageous to
construct asymmetries which vanish as CP is restored.

CP violation will be realized if at least one of the CP-even terms is
present (i.e.\ either $a \neq 0$ and/or $b \neq 0$) and $c$ is
non-zero. In the following we keep the three coefficients non-zero in
our analytical work, where appropriate. However, in the numerical
presentation of most of our results we will take $b=0$ for simplicity,
keeping non-zero $b$ only where essential.  Further, we make the
justified approximation to neglect the possible momentum dependence of
the form factors.

Notice that neither $q_{1\mu} V_{HZZ}^{\mu \nu}$ nor $q_{2\nu}
V_{HZZ}^{\mu \nu}$ are zero, i.e.\ the Ward identities are
violated. This is due to the breaking of electroweak symmetry and is
already the case for the SM vertex. Some studies, e.g.\
Refs.~\cite{Biswal:2005fh,Allanach:2006yt}, explicitly construct the
extra terms such that they satisfy such Ward identities individually,
for example, by taking a CP-even term of the form $q_1 \cdot q_2 \,
g_{\mu\nu} - q_{2\mu} q_{1\nu}$. Strictly speaking, this is not
necessary as long as any additional terms vanish in the limit $m_Z \to
0$. Furthermore, since one must separately include the SM $g^{\mu
\nu}$ coupling and the new CP-even contribution (with independent
coefficients), one may always reproduce our choice of the vertex with a
suitable redefinition of the coefficients.

Our vertex differs from the vertex of
Refs.~\cite{Miller:2001bi,Choi:2002jk} only in the choice of the
normalisation of the coefficients (to make them dimensionless).  The
normalisation of the coefficients (and the overall normalisation) also
differs from Refs.~\cite{Buszello:2002uu,Buszello:2006hf}, where $m_H$
was used in contrast to our $m_Z$. Additionally,
Refs.~\cite{Buszello:2002uu,Buszello:2006hf} use the momenta of the
Z-bosons to define the last term (i.e.\ $\sim
q_1^{\alpha}q_2^{\beta}$) in contrast to our $\sim
p^{\alpha}k^{\beta}$. However, this last difference is for this
process only a factor of $-2$ since the additional terms
are removed by the asymmetric property of the tensor. Finally,
Ref.~\cite{Biswal:2005fh} differs in the choice of the last term (again
$\sim q_1^{\alpha}q_2^{\beta}$) and rearranges the contributions of
the first two terms, as discussed in the preceding paragraph. For
$b=0$ and light lepton final states, all these vertices are the same,
modulo momentum independent normalisations of the coefficients.  

From the above discussion it is clear that the total decay rate of
Eq.~(\ref{hdecayzz}), which is CP-even and $\tilde{\rm T}$ even, can only
probe $a$, $\Re e(b)$ and the absolute values of $b$ and $c$. In order to 
probe the other non-standard parts of the $H ZZ$ coupling, in particular in
order to probe CP-violation, one must construct observables that are
odd under CP and/or $\tilde {\rm T}$. These observables
give rise to various azimuthal and polar asymmetries and will make
their presence felt through rates which are integrated over a partial
(non-symmetric) phase space. Thus one may probe $\Re e(b)$, $\Im m(b)$, 
$\Re e(c)$ and $\Im m(c)$ either by using the shapes of various kinematical
distributions or by constructing observables which are obtained using
partially integrated cross 
sections~\cite{Chang:1993jy,Han:2000mi,Biswal:2005fh}\footnote{In fact, 
Ref.~\cite{Biswal:2005fh} constructed systematically the whole
set of asymmetries which probe different parts of the anomalous couplings.}.
We will use the latter to construct asymmetries which receive contributions 
from non-standard couplings and which vanish in the tree-level
SM. These are related to simple counting experiments, recording the
number of events in well defined regions of the phase space.
It may also be noted that results obtained using these asymmetries are
less sensitive to the effect of radiative corrections to the 
production~\cite{nloggcorrs} and decay~\cite{fleischer,dennereal} of the Higgs boson.

\begin{figure}[tbh]
\begin{center}
\includegraphics[width=14cm,clip=true]{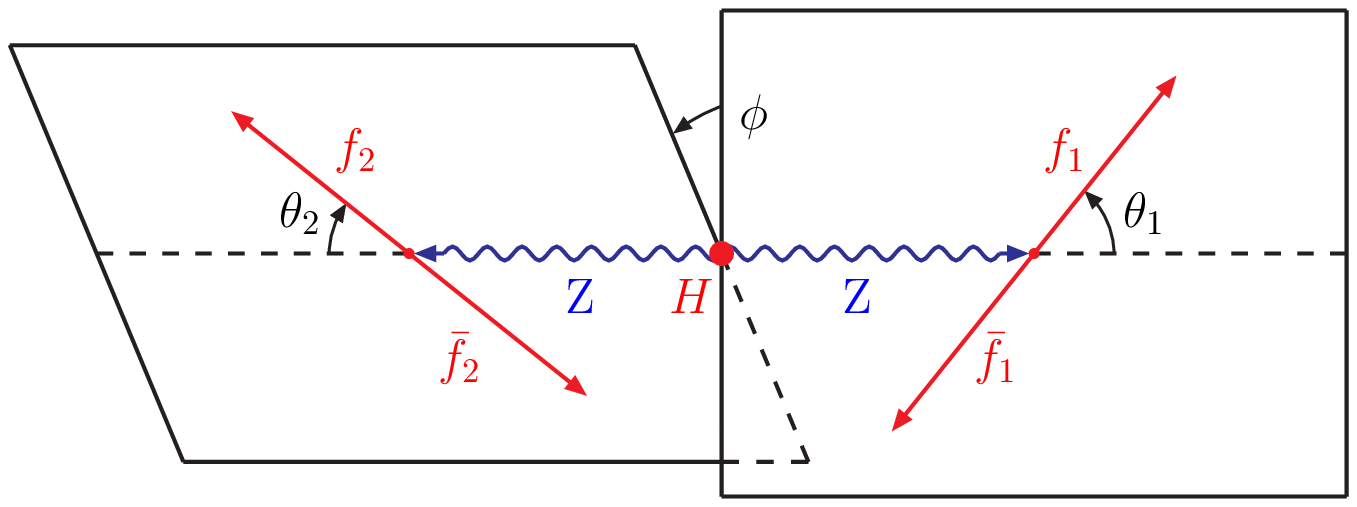}
\caption{The definition of the polar angles ${\theta_i}$ ($i=1,2$) and
the azimuthal angle $\phi$ for the sequential decay \mbox{$H
\rightarrow ZZ^{(*)} \rightarrow (f_1\bar{f}_1) \, (f_2\bar{f}_2)$}.}
\label{gmmm_angles}
\end{center}
\vspace*{-0.3cm}
\end{figure}

In order to find observables which project out the various
non-standard couplings in Eq.~(\ref{param}) it is instructive to have
an analytical formula for the differential distribution of the Higgs
decay to off-shell $Z$ bosons with subsequent decay into fermion pairs
with respect to the various scattering angles. We denote the polar
angles of the fermions $f_1,f_2$ in the rest frame of the parent $Z$
bosons by $\theta_1$ and $\theta_2$, and the azimuthal angle between
the planes formed from the fermion pairs in the Higgs rest frame by
$\phi$ [see Fig.~\ref{gmmm_angles}].  Also note that there can be no
angular correlations (at tree-level) between the initial and final
states (i.e. between the beam-direction and the final state leptons)
as long as the Higgs has zero spin.

Introducing the notation $c_{\theta_i}\equiv \cos\theta_i$,
$s_{\theta_i}\equiv\sin\theta_i$ ($i=1,2$), $c_\phi\equiv \cos\phi$,
etc.\ the tree-level differential decay rate for distinguishable
fermions can be cast into the form
\br
\frac{d^3\Gamma}{dc_{\ta}\, dc_{\tb} d\phi}
&\sim&
a^2 \left[ 
s_{\ta}^2 s_{\tb}^2 
- \frac{1}{2\gamma_a} s_{2\ta} s_{2\tb} \cph 
+ \frac{1}{2\gamma_a^2}
\left[
(1+c_{\ta}^2)(1+c_{\tb}^2)
+s_{\ta}^2 s_{\tb}^2 c_{2\phi} \right] \right. \nonumber\\
&& \phantom{a^2 \Big[} \left.
-\frac{2\eta_1 \eta_2}{\gamma_a} 
\left(s_{\ta} s_{\tb} \cph - \frac{1}{\gamma_a} c_{\ta} c_{\tb} \right) \right]
\nonumber \\
&+& |b|^2 \frac{\gamma_b^4}{\gamma_a^2} \, x^2
\, s_{\ta}^2 s_{\tb}^2 \nonumber \\
&+& 
|c|^2 \frac{\gamma_b^2}{\gamma_a^2} \, 4 x^2 \,
\left[ 1 + c_{\ta}^2 c_{\tb}^2 - \frac{1}{2} s_{\ta}^2 s_{\tb}^2(1 + 
c_{2\phi}) +2\eta_1\eta_2 c_{\ta} c_{\tb} \right] \nonumber \\
&-& 2a\Im m(b)
\frac{\gamma_b^2}{\gamma_a^2} \, x \, s_{\ta} s_{\tb}
s_\phi \left[\eta_2 c_{\ta} +\eta_1 c_{\tb} \right]
\nonumber \\
&-& 2a\Re e(b) \frac{\gamma_b^2}{\gamma_a^2} \, x
\, \left[-\gamma_a s_{\ta}^2 s_{\tb}^2 + \frac{1}{4}s_{2\ta}s_{2\tb} \cph 
+\eta_1 \eta_2 s_{\ta} s_{\tb}\cph \right] \nonumber \\
&-& 2a\Im m(c) \frac{\gamma_b}{\gamma_a} 
\, 2 x \, 
\left[ \lefteqn{\phantom{\frac{1}{\gamma_a}}} -s_{\ta} s_{\tb} \cph (\eta_1 c_{\tb}+\eta_2 c_{\ta}) \right.
\nonumber \\ & & \qquad \qquad \qquad \qquad
\left. + \frac{1}{\gamma_a}  \left( \eta_1 c_{\ta} (1+c_{\tb}^2) + \eta_2 c_{\tb} (1+c_{\ta}^2)
\right) \right] \nonumber\\
&-& 2a\Re e(c) \frac{\gamma_b}{\gamma_a} \, 2x \,
s_{\ta} s_{\tb} s_{\phi} 
\left[ -c_{\ta} c_{\tb} + \frac{s_{\ta} s_{\tb} \cph}{\gamma_a} -\eta_1 \eta_2 \right] 
\nonumber \\ 
\nonumber\\
&+& 2\Im m(b^* c) 
\frac{\gamma_b^3}{\gamma_a^2} \, 2x^2 \,
s_{\ta} s_{\tb} \cph \left[\eta_2 c_{\ta} + \eta_1 c_{\tb} \right] 
\nonumber\\
&+ & 
2\Re e(b^* c) 
\frac{\gamma_b^3}{\gamma_a^2}\, 2x^2 \, s_{\ta} s_{\tb} s_\phi 
\left[ c_{\ta} c_{\tb} + \eta_1 \eta_2\right] \;,
\label{mmstar}
\er
where $x= m_1 m_2/m_Z^2$ with $m_1,m_2$ the virtualities of the $Z$ bosons
($q_i^2=m_i^2$).  Furthermore, we have introduced the
notation $\gamma_a = \gamma_1 \gamma_2 (1+\beta_1\beta_2)$ and
$\gamma_b = \gamma_1 \gamma_2 (\beta_1+\beta_2)$ in terms of the
Lorentz boost factors of the $Z$ bosons, $\gamma_{i} =
1/\sqrt{1-\beta_i^2}$, and the velocities \br \beta_{i} =
\frac{m_H}{2\,E_{i}} \beta \qquad i=1,2 \;, \er where $E_{i}$ are
the $Z$ boson energies in the Higgs rest frame and 
\be \beta = \left\{ \left[1-\frac{(m_1+m_2)^2}{m_H^2}\right]
\left[1 - \frac{(m_1-m_2)^2}{m_H^2}\right] \right\}^{1/2}. \ee
The $\eta_i$ are given in terms of the weak vector and axial couplings $v_{f_i},\,a_{f_i}$,
\be \eta_i = \frac{2\,v_{f_i}a_{f_i}}{v_{f_i}^2+a_{f_i}^2}, 
\qquad \mbox{with}
\qquad v_{f_i} = T^3_{f_i} - 2  Q_{f_i} \sin^2\theta_W, 
\qquad a_{f_i} = T^3_{f_i}\;.  \ee
Here $T^3_{f_i}$ denotes the third component of the weak isospin and 
$Q_{f_i}$ the electric charge of the fermion $f_i$, in our case
$e^-$ or $\mu^-$.\\[0.2cm]
%
%
\section{Sensitivity of the total production to new couplings}
\label{sectthree}
As discussed in section \ref{secttwo}, one may use the total decay rate
of the process in Eq.~(\ref{hdecayzz}) to test possible deviations
from the SM in the Higgs to $ZZ$ coupling. At the LHC the
dominant Higgs production process is given by gluon--gluon fusion, 
\br 
gg \to H \to ZZ^{(*)} \to (f_1\bar{f_1}) (f_2\bar{f_2}) \;,
\label{ggrate} 
\er
with $f=e$ or $\mu$. The width for the process $H \to Z Z^{(*)}
\to (f_1\bar{f_1}) \, (f_2\bar{f_2})$ is given by,
\br
\lefteqn{\Gamma (H \to Z Z^{(*)}
\to (f_1\bar{f_1}) \, (f_2\bar{f_2}))=} && \nonumber\\
&& 
\frac{1}{\pi^2} 
\int_0^{m_H^2}       \!\!\! dm_1^2
\int_0^{[m_H-m_1]^2} \!\!\! dm_2^2
\frac{m_Z \, \Gamma_{Z \to f_1\bar{f_1}}}{[(m_1^2-m_Z^2)^2 + m_Z^2 \Gamma_Z^2]}
\frac{m_Z \, \Gamma_{Z \to f_2\bar{f_2}}}{[(m_2^2-m_Z^2)^2 + m_Z^2 \Gamma_Z^2]}
\, \Gamma_{H \to Z Z} \, , 
\label{htozzto4f}
\er
where the width for the Higgs decay to two Z bosons\footnote{For the
on-shell decay $H \to ZZ$, see Ref.~\cite{Cahn:1988ru}.} of virtualities
$m_1$ and $m_2$ is,
\br
\Gamma_{H \to ZZ}
&=& \frac{G_F m_H^3}{16\sqrt{2}\pi}\, \beta \, 
\left\{ a^2 \left[\beta^2+\frac{12m_1^2 m_2^2}{m_H^4} 
\right] + 
|b|^2 \frac{m_H^4}{m_Z^4} \, \frac{\beta^4}{4} 
+ |c|^2 x^2 \, 8\beta^2 \right.
\nonumber\\
&& \qquad \qquad \qquad \left. +
a\Re e(b) \frac{m_H^2}{m_Z^2} \, \beta^2 \, 
\sqrt{\beta^2+4 m_1^2 m_2^2/m_H^4} 
\right\}
\label{htoz1z2}
\er
and $\Gamma_{Z \to f_i \bar f_i}$ is the width for the decay of a
$Z$ boson to a fermion pair, $f_i \bar f_i$, as given in the SM,
\begin{equation}
\Gamma_{Z \to f_i \bar f_i} = \frac{G_F m_Z^2}{6 \sqrt{2} \pi} \, m_Z \, 
(v_{f_i}^2+a_{f_i}^2) \;.
\end{equation}

As expected, the CP$\tilde{\rm T}$-even total rate cannot directly test
CPV (since there is no interference between the CP-even and CP-odd
terms), but it is sensitive to possible non-SM coupling effects in
$\Re e(b)$ and the absolute values of $b$ and $c$. Furthermore,
Eq.~(\ref{htoz1z2}) shows that the linear rise in $\beta$ just below
the threshold is typical \cite{Miller:2001bi} of the SM Higgs
boson\footnote{This observation is valid for all spins, with one minor
caveat: the spin-2 case can also have a term which presents a linear
rise in $\beta$ but this can be excluded by angular correlations, see
Ref.~\cite{Miller:2001bi}.}.

The Tevatron is in principle also sensitive to the process of
Eq.~(\ref{ggrate}) for sufficiently high Higgs boson masses. Indeed,
preliminary Tevatron results~\cite{tev_prelim} indicate that a signal
for a Higgs boson of $150\,{\rm GeV}$ would have been seen (with
$95\%$ confidence) if the observed(expected) D0-CDF combined total
cross-section were enhanced by a factor of $2.4(3.3)$. However, this
result is dominated by the decay $H \to W^+W^-$; the $H \to ZZ$ decay
is suppressed relative to $W^+W^-$ by around a factor of $10$ for a
$150\,{\rm GeV}$ Higgs boson, so an enhancement of the $HZZ$ vertex
from additional couplings would need to be very large indeed to be
seen by the Tevatron. Since we are here investigating the $HZZ$
coupling, we make the assumption that the other decay channels are
unaffected and that any change originates from the $HZZ$ coupling
alone. For lower Higgs masses the $HZZ$ coupling can also play a role
in the production of the Higgs via the channel $q \bar q \rightarrow
Z^* \rightarrow Z H$. However, as can be seen from
Ref.~\cite{Gonzalez:2007zza}, with current data, the Tevatron would
be sensitive to this production mode only if the cross-section were
enhanced by a factor of $\sim 30-90$ compared to the SM and thus the
nonobservation of this channel in the current data only puts very weak
constraints on the magnitude of these couplings.

To estimate the sensitivity of the LHC to deviations from the SM coupling,
we refer to the ATLAS study for the process of Eq.~(\ref{ggrate}) at
$m_H = 150\,$GeV and $200\,$GeV~\cite{tdratl,Hohl:2001atl}.  In this
study, four leptons were selected using the standard electron and muon
identification criteria. Events were required to have two leptons with
$p_T>20\,$GeV and two additional leptons with $p_T>7\,$GeV, with
rapidity $|\eta|<2.5$ for all four. The signal and background were
compared in a small mass window around the Higgs boson mass, and a lepton
identification and reconstruction efficiency was applied.

For the $m_H=150\,$GeV analysis, one lepton pair was required to have
an invariant mass within $10\,$GeV of $m_Z$ while the other pair was
required to have an invariant mass above $30\,$GeV. Additionally,
isolation and impact parameter cuts were used to further remove
irreducible backgrounds. For the $200\,$GeV analysis, the continuum
$ZZ$ background was further removed by requiring the $p_T$ of the
hardest $Z$-boson to be greater than \mbox{$m_H/3 \approx 66.6\,$GeV} (see
Refs.~\cite{tdratl,Hohl:2001atl} for further details).

Note that this ATLAS study was performed at tree-level with no
K-factors.  Higher order corrections to the production process could
alter the cross section by up to a factor two \cite{nloggcorrs}. The
higher order electroweak corrections to the Higgs decays into $W/Z$
bosons have been calculated in Ref.~\cite{fleischer} in the narrow
width approximation. Ref.~\cite{dennereal} presents the complete
${\cal O}(\alpha)$ corrections to the general $H\to 4l$ processes,
including off-shell gauge bosons which are important
for our study. The corrections have been shown to change the partial
width by up to 5\% for the Higgs boson masses we consider in this
paper. Our analysis, which uses the results of the ATLAS study,
strictly speaking is only valid at tree-level, despite the all-orders
validity of the $HZZ$ coupling (see section~\ref{secttwo}).

After these cuts, the study found for a $150\,$GeV Higgs boson and an
integrated luminosity of 100~fb$^{-1}$, $67.6$ signal events with a
background of $8.92$ events. The corresponding signal and background
events for a $200\,$GeV Higgs boson were $54$ and $7$, for an
integrated luminosity of $30\, {\rm fb}^{-1}$. Altered $HZZ$ couplings
will enhance (or decrease) the number of signal events, while leaving
the number of background events fixed. However, the size of this
enhancement (or reduction) is model-dependent.  Although the change in
the {\rm width} for $H \to ZZ^* \to 4l$ is clear from
Eq.~(\ref{htoz1z2}), the {\it branching ratio} depends on how the
other Higgs decay channels are affected by the new physics. As
mentioned above, we here make the assumption that only the $HZZ$ vertex
deviates from that of the SM.  If this were not the case, and, for
example, the $HWW$ coupling was similarly enhanced, then any
enhancement of the $H \to ZZ$ branching ratio would be watered
down. Furthermore, we assume the Higgs production proceeds as
according to the SM, since the dominant production mode contains no
$HZZ$ coupling, but one should be aware that CPV effects in other
vertices may alter the Higgs production rate (see e.g.\
Ref.~\cite{Dedes:1999zh}).  Finally, we assume that the rate
calculated with the general $H ZZ$ coupling Eq.~(\ref{param}) will be
reduced by experimental cuts in the same way as the \sm rate. Only
electron and muon final states are considered, and we scale up the
number of signal and background events to correspond to an integrated
luminosity of $300\, {\rm fb}^{-1}$.

\begin{figure}[tbh]
\begin{center}
\includegraphics[width=7cm,clip=true]{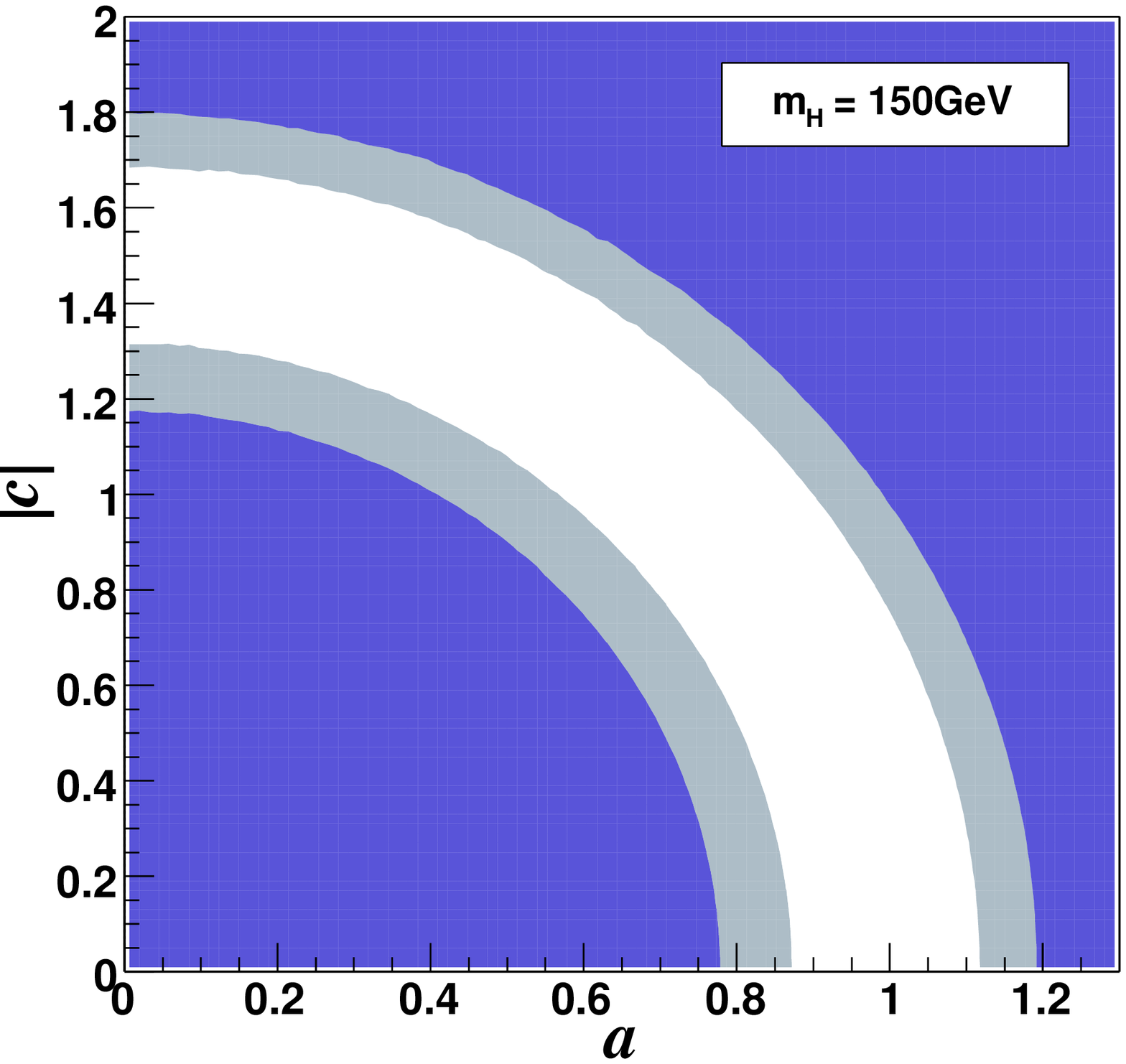}
\includegraphics[width=7cm,clip=true]{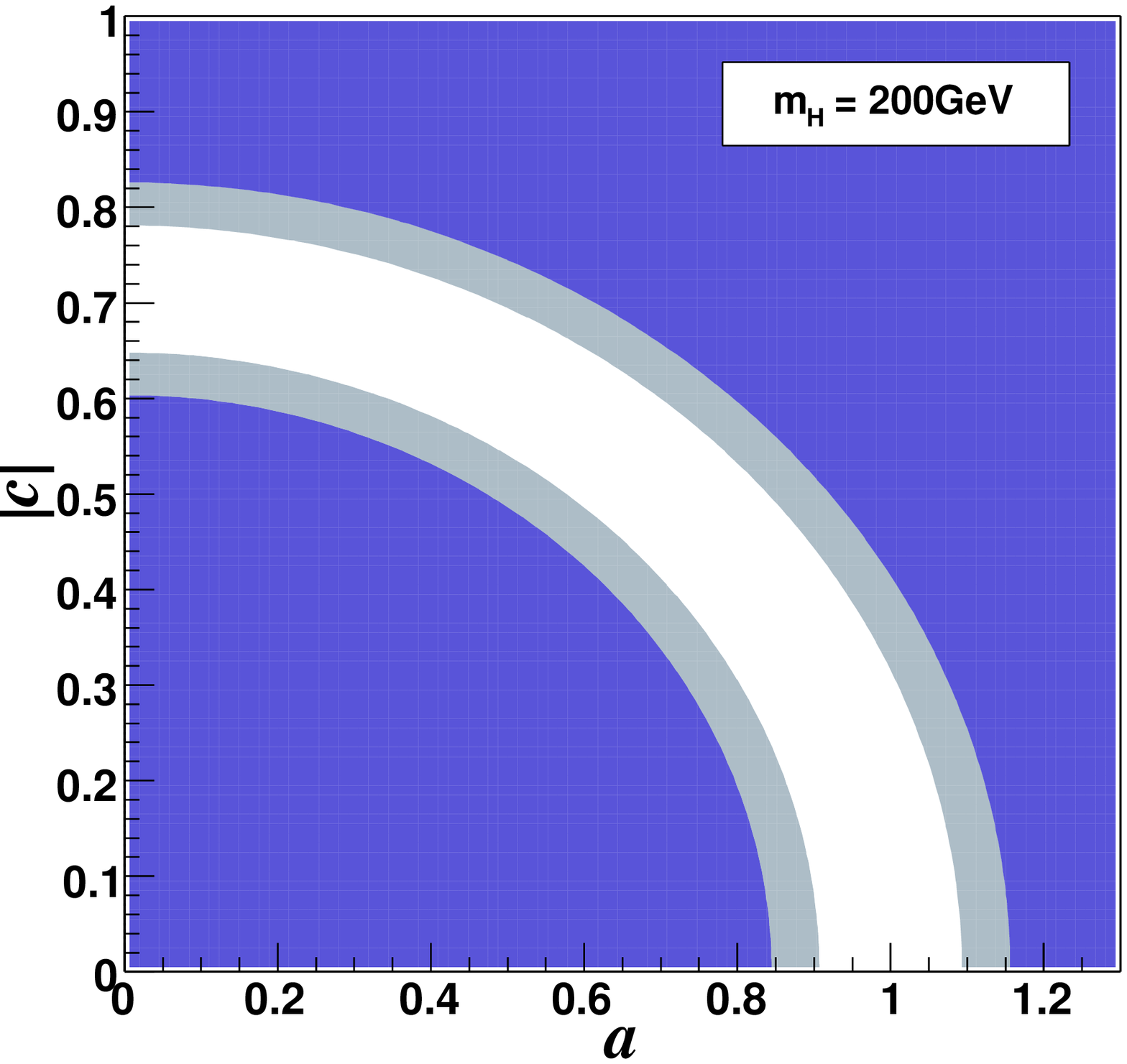}
\caption{The number of standard deviations from the SM which can be
obtained in the process $gg\to H\to Z^* Z^* \to 4$~{\it leptons}, as a
scan over the $(a,|c|)$ plane. The Higgs mass has been chosen to be
$150\,$GeV {\it (left)} and $200\,$GeV {\it (right)}. The white region
is where the deviation from the SM is less than $3 \, \sigma$; in the
light blue/light grey region the deviation is between $3\,\sigma$ and
$5\,\sigma$; while for the dark blue/dark grey region the deviation is
greater than $5\,\sigma$.}
\label{contourplot}
\end{center}
\vspace*{-0.3cm}
\end{figure}

We then calculate the total number of signal events $N_S$ that we
expect from the new coupling and compare the expected change (with
respect to the SM) with the possible statistical fluctuations of the
\sm signal and backgrounds. The significance of this deviation from
the SM expectation (in units of one standard deviation) is then
$(N_S-N_S^{\rm SM})/\sqrt{N_S^{\rm SM} + N_B}$, where $N_S^{\rm SM}$
is the number of signal events expected in the \sm and $N_B$ is the
number of background events. This quantity, for $m_H=150\,$GeV and
$200\,$GeV, is plotted in Fig.~\ref{contourplot}, where we have
scanned over values of the couplings $a$ and $|c|$ (the total rate is
independent of the phase of $c$). For simplicity, we have set $b=0$
(for $b=0$, one can see that Eq.~(\ref{htoz1z2}) is symmetric in $a$
allowing us to restrict the plot to positive values).  As can be
inferred from Fig.~\ref{contourplot} in the white region we can not
distinguish the corresponding $a,\,c$ values from the SM case,
$a=1,\,c=0$ at a significance more than $3 \,\sigma$.

Large values of $|c|$, however, together with the SM value of $a=1$
are easily identified at the LHC. For example, the scenario $a=c=1$ is
excluded with around $5\,\sigma$ significance for $m_H=150\,$GeV and
over $20 \, \sigma$ significance at $m_H=200\,$GeV. However, since
$|c|$ arises from new physics one would expect its value to be
suppressed by the size of the new physics scale, and therefore be
rather small. For $a=1$ (the SM value) we find that this measurement
provides $3\,\sigma$ evidence of non-zero $c$ only if $c\gtrsim 0.75$
or $c\gtrsim 0.32$, for $m_H=150\,$GeV and $200\,$GeV,
respectively. Furthermore, since both $a^2$ and $|c|^2$ contribute to
the total rate, we cannot distinguish whether or not any deviation is
originating from non-standard values of $a$ or $|c|$, and even if the
SM total rate is confirmed, one cannot definitively say that $a$ and
$c$ take their SM values since an enhancement in $|c|$ may be
compensated by a reduction in $a$. Also, a non-zero value of $b$ could
provoke a similar effect. Indeed, the total rate is not even reliable
in distinguishing a CP-even eigenstate from a CP-odd one. Instead, to
provide a definitive measurement of CP violation in this coupling, one
must explore asymmetries which probe the interference of the CP-even
and CP-odd contributions directly.

\section{Asymmetries as a probe of CP-violation}
\label{sectfour}

As stated above, apart from the terms proportional to $a$ and $\Re
e(b)$, all other contributions to the vertex Eq.~(\ref{param}) are odd
under CP and/or $\tilde{\rm T}$ transformations, and their presence implies
violations of the corresponding symmetries in the interaction. We
exploit this by constructing observables from the 3-momenta of the
initial and final state particles with the {\it same} transformation
property under the discrete symmetries as one of these non-SM
couplings. The expectation value of the {\it sign} of such a variable
will directly probe the corresponding coupling
coefficient \cite{Biswal:2005fh}.\footnote{This statement is true 
strictly when only the linear terms in the anomalous $H ZZ$ coupling 
are kept. Potentially, the asymmetries may  also contain combinations of 
more than one (small) anomalous couplings which will have the same discrete 
symmetry transformation properties. In that case the asymmetry will be a 
direct probe of that particular combination of the non-SM couplings.} 
The asymmetry will be proportional to the probed coupling and therefore 
non-zero only if the corresponding non-SM coupling is present. Furthermore, 
since these asymmetries are exactly zero for all backgrounds (we neglect 
interference effects), backgrounds cannot contribute to the asymmetry, except 
through fluctuations, and  it is therefore possible to use less stringent 
cuts on the signal.

In this section we present various observables and their
asymmetries which allow one to probe the real and imaginary parts of the
form factors $b$ and $c$, the latter being indicative of CP violation
for simultaneously non-zero $a$ and/or $b$ values. \\[0.3cm]
%
\noindent {\bf 1. An observable to probe \boldmath{$\Im m(c)$}:} 
\hspace*{0.2cm}
We consider the observable
\be
O_1 \equiv \frac{(\vec{p}_{2Z} - \vec{p}_{1Z}) \cdot
(\vec{p}_{3H} + \vec{p}_{4H})}{|\vec{p}_{2Z} - \vec{p}_{1Z}|
|\vec{p}_{3H} + \vec{p}_{4H}|} \;.
\label{o1}
\ee 
Here $\vec{p}_i$, $i=1, \ldots 4$ are the 3-momenta of the leptons
(in the order $f_1 \bar f_1 f_2 \bar f_2$), and the subscripts $Z$ and $H$
denote that the corresponding 3-vector is taken in the $Z$ boson or
Higgs boson rest frame, respectively.  This observable is CP odd and
$\tilde{\rm T}$ even and thus probes the non-SM coupling with the same
transformation properties, {\it i.e.} $\Im m(c)$.  With the above
angular definitions we have
\br
O_1 = \cos \theta_1 \; .
\label{o1theta1}
\er
We can calculate the resulting asymmetry by integrating
Eq.~(\ref{mmstar}) over the angles with an appropriate
weighting. Although Eq.~(\ref{mmstar}) is only valid for
distinguishable fermions, we may include fermions of the same flavour,
e.g.\ $(e^-e^+)(e^-e^+)$, and distinguish the fermions by the
requirement that the first pair reconstruct the $Z$-boson mass. In
general, the contribution from the same final state with the
antiparticles switched would contain two off-shell $Z$-bosons and may
be neglected. However, one should also note that this observable
requires one to distinguish between fermions and anti-fermions.

\begin{figure}[tbh]
\begin{center}
\includegraphics[width=8cm,clip=true]{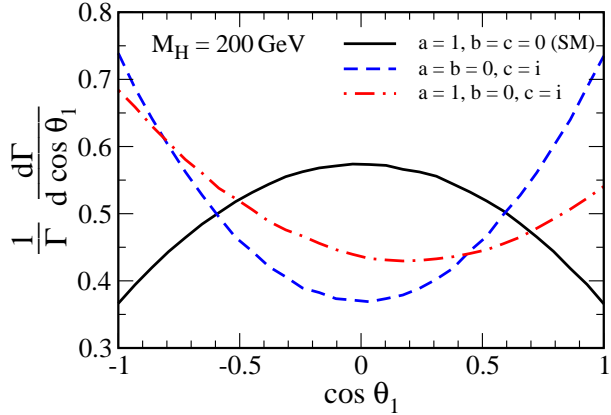}
\end{center}
\caption{The normalized differential width for \mbox{$H \rightarrow Z
Z \rightarrow (f_1\bar{f}_1)\,(f_2\bar{f}_2)$} and \mbox{$m_H=200\,$~GeV} with
respect to the cosine of the fermion $f_1$'s polar angle $\theta_1$. The 
solid (black) curve shows the \sm case ($a=1$, $b=c=0$) while the dashed 
(blue) curve is for a pure CP-odd state ($a=b=0$, $c=i$). The dot-dashed 
(red) curve is for a state with a CP violating coupling ($a=1$, $b=0$, $c=i$). 
One can clearly see an asymmetry about $\cos \theta_1=0$ for the CP
violating case.}
\label{gmm_theta}
\end{figure}

The angular distribution of Eq.~(\ref{mmstar}) contains several terms
linear in $\cos \theta_1$. However, most of these terms are removed by
integration over the angles $\theta_2$ and $\phi$, leaving only one term
proportional to $a \,\Im m(c)$. So only a non-zero value of $\Im m (c)$
gives rise to this forward-backward asymmetry and hence provides a definitive
signal of CP violation in the $H ZZ$ vertex. This is demonstrated in
Fig.~\ref{gmm_theta}, which shows the dependence on $\cos \theta_1$
for pure CP-even, pure CP-odd and CP-violating
interactions\footnote{This figure differs from the corresponding
figure in Ref.~\cite{Allanach:2006yt} for the CP violating coupling due to the
different conventions. The corresponding curve for the mixed CP state in 
Ref.~\cite{Allanach:2006yt} is reproduced with our current conventions if 
$a=1, b=0, c=-i/2$.}.

To quantify the effect we define an asymmetry by,
\br
{\cal A}_1 = \frac{\Gamma (\cos\theta_1 > 0)-\Gamma (\cos\theta_1<0)}
{\Gamma (\cos\theta_1 > 0)+\Gamma (\cos\theta_1<0)} \;.
\label{asymm1def}
\er
This asymmetry, which is the expectation value of the sign of $\cos\theta_1$
(Eq.~\ref{o1theta1}) and which is CP-odd and $\tilde{\rm T}$ even, directly probes
$\Im m(c)$ which is also CP-odd and $\tilde{\rm T}$ even.
Integrating Eq.~(\ref{mmstar}), the asymmetry ${\cal A}_1$ can be written as
\br
{\cal A}_1 &=& \frac{1}{\tilde \Gamma} \, 
\int d^2{\cal P} \, \beta \, \left\{ -3\, a \Im m(c) x\,
\eta_1 \, \gamma_b \right\},
\label{gmm_asym1_eq}
\er
where $\tilde \Gamma$ is related to the decay width $H\to ZZ^{(*)} \to (f_1 
\bar f_1) \, (f_2 \bar f_2)$, {\it c.f.} Eqs.~(\ref{htozzto4f},\ref{htoz1z2}), 
and is given by
\begin{equation}
\tilde \Gamma =
\int d^2{\cal P} \, \beta \,
\left\{ a^2  \left(1 + \frac{\gamma_a^2}{2} \right)
    + |b|^2  \frac{\gamma_b^4}{2} x^2
                  + 4|c|^2 x^2 \gamma_b^2
                  + a \Re e(b) x \gamma_a \gamma_b^2 \right\},
\end{equation}
and the integral is over the virtualities, weighted with the Breit-Wigner 
form of the Z-boson propagators,
\begin{equation}
\int d^2{\cal P} \, \cdots =
\int_0^{m_H^2}       \!\!\! dm_1^2
\int_0^{[m_H-m_1]^2} \!\!\! dm_2^2 \,
\frac{m_1^2}{[(m_1^2-m_Z^2)^2+m_Z^2\Gamma_Z^2]}
\frac{m_2^2}{[(m_2^2-m_Z^2)^2+m_Z^2\Gamma_Z^2]} \ldots.
\label{prop}
\end{equation}
This asymmetry is calculated at tree-level. Higher order electroweak
corrections to the decay $H \to ZZ \to 4$~leptons are of the order
5-10\% for angular distributions~\cite{fleischer,dennereal}. One might
worry that these corrections could feed into the asymmetry and swamp
the signal. However, unless the corrections introduce some new effect
(and are thus in some sense ``leading order''), one expects their
contribution to CP violation to be of a similar proportion as those at
tree-level, so they would provide a correction to
Eq.~(\ref{gmm_asym1_eq}) of 5-10\%, and not significantly alter our
results.

\begin{figure}[tbh]
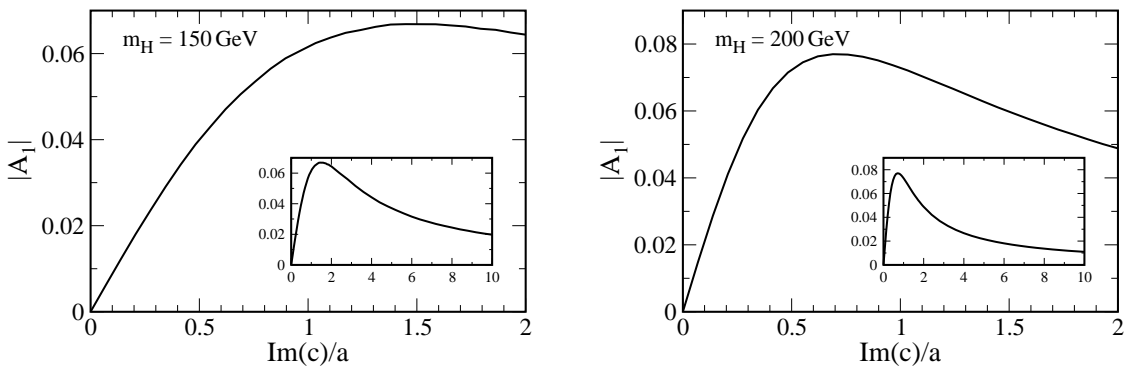

\begin{center}
\includegraphics[width=7cm,clip=true]{a1_mag_150.eps} \hspace{6mm}
\includegraphics[width=7cm,clip=true]{a1_mag_200.eps}
\caption{The asymmetry ${\cal A}_1$ given by Eq.~(\ref{gmm_asym1_eq}) as a
function of the ratio $\Im m(c)/a$, for a Higgs boson of mass
$150\,$GeV ({\it left}) and 200~GeV ({\it right}). We chose $b=0$.
The inserts show the same quantities for a larger range of $\Im m(c)/a$.}
\end{center}
\label{gmm_asym1}
\vspace*{-0.3cm}
\end{figure}

Fig.~\ref{gmm_asym1} shows the values of ${\cal A}_1$ for a Higgs mass
of 150 and 200 GeV, respectively, as a function of the ratio $\Im m(c)/a$ and 
where we have set $b =0$ for simplicity.  
The value $\Im m(c)/a=0$ corresponds to the purely scalar state and
$\Im m(c)/a \to \infty$ to the purely CP-odd case. It is clear from
Eq.~(\ref{gmm_asym1_eq}) that ${\cal A}_1$ is sensitive only to the
relative size of the couplings since any overall factor will cancel in the
ratio, {\it c.f.} Eq.~(\ref{asymm1def}). We find that the asymmetry is 
maximal for $\Im m(c)/a \sim
1.5(0.7)$ with a value of about $0.067(0.077)$ for
$m_H=150(200)\,$GeV. The smallness of this asymmetry arises
from the fact that it is proportional to the coupling
$\eta_1=2v_1a_1/(v_1^2+a_1^2)$ which is equal to approximately $0.149$ for
$e,\mu$ final states, {\it c.f.}  Eq.~(\ref{gmm_asym1_eq}). 

In order to estimate whether this asymmetry can be measured at the
LHC, we calculate the significance with which a particular CP
violating coupling would manifest. To do this, we must take into
account the backgrounds to the signal process, which will contaminate
the asymmetry in two ways. Firstly, despite being CP-conserving the
backgrounds may contribute to the numerator of the asymmetry via
statistical fluctuations (e.g.\ the background events with $O_1>0$ may 
fluctuate upwards while those with $O_1<0$ may fluctuate
downwards and vice versa). Secondly, they will directly contribute to the 
denominator of the asymmetry. 

Consequently, the measured asymmetry will be given by,
\begin{equation}
{\cal A}_1^{\rm meas} = \frac{N_S^{\rm asym}}{N_S + N_B}
= {\cal A}_1 \frac{N_S}{N_S + N_B},
\end{equation}
where $N_S^{\rm asym}$ is the asymmetry in the number of events in the two
hemispheres, and ${\cal A}_1$ is the perfect theoretical asymmetry given in
Eq.~(\ref{asymm1def}).

The statistical fluctuation in an asymmetry calculated using a total number of
events  $N = N_B \,+\, N_S$, even when $N_B$ and $N_S$ are expected to be
symmetric, is $1 /\sqrt{N}$. Hence, the significance of the expected 
asymmetry, $S$, in units of this statistical fluctuation is given by
\begin{equation}
S = 
{\cal A}_1^{\rm meas}  {\sqrt{N}} = \frac{N_S^{\rm asym}}{\sqrt{N}} = {\cal A}_1 \frac{N_S}{\sqrt{N}}\;.
\end{equation}
In order to calculate this, we need to know the number of signal and
background events expected at the LHC. However, in this case, since
the contamination of the significance from the background is rather
minimal, we choose to use the event sample {\it before} the detailed
cuts to remove backgrounds, but after the initial selection cuts. For 
$150\,$GeV we take the number of
signal and background events before applying the additional isolation
and impact parameter cuts to remove the irreducible backgrounds, and
for $150\,$GeV we do not apply the final $p_T$ cut on the hardest
$Z$-boson (see Refs.\cite{tdratl,Hohl:2001atl}).

Then, according to Refs.\cite{tdratl,Hohl:2001atl}, for a
\mbox{$m_H=150\,$GeV} SM Higgs boson, we have a signal cross-section
of $5.53\,$fb, with an overall lepton efficiency of
$0.7625$. Assuming an integrated luminosity of $300\,{\rm fb}^{-1}$
this gives $1265$ signal events. For \mbox{$m_H=200\,$GeV}, the
corresponding signal is $1340$ events.  The number of signal events
for the CP violating case is then obtained by multiplying the number
of \sm events by the ratio of CP violating to SM branching ratios. In
the CP-violating case we always assume the SM value for the CP-even
coefficient, $a=1$.  For simplicity we assume the charge of the
particles to be unambiguously determined, and pair the leptons by
requiring at least one pair to reconstruct the $Z$ boson mass. The
number of background events before cuts has been derived
correspondingly from the study Refs.\cite{tdratl,Hohl:2001atl} and
amounts to $1031(740)$ events for $m_H=150(200)\,$GeV.

\begin{figure}[tbh]
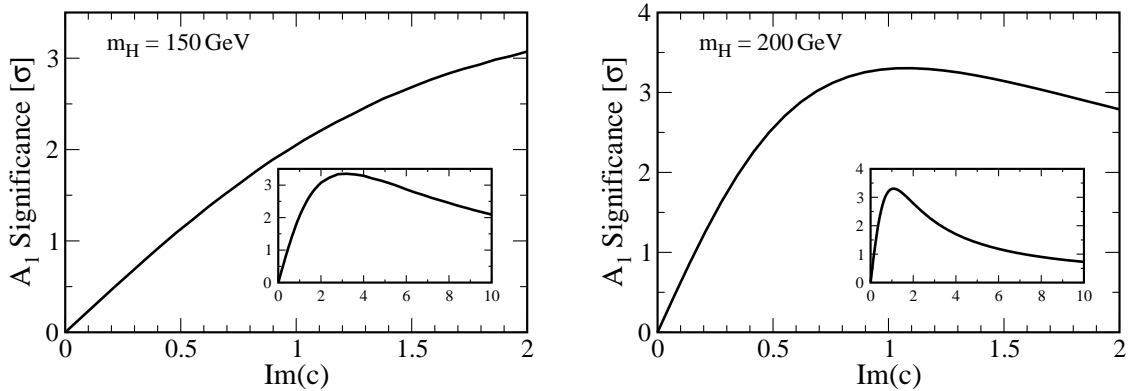

\begin{center}
\includegraphics[width=7cm,clip=true]{a1_sig_150.eps} \hspace{6mm}
\includegraphics[width=7cm,clip=true]{a1_sig_200.eps}
\caption{The significances corresponding to the asymmetry ${\cal A}_1$ 
as a function of $\Im m(c)$, for a Higgs boson of mass
$150\,$GeV ({\it left}) and 200~GeV ({\it right}). We chose the CP-even
coupling coefficient $a=1$ and $b=0$. The inserts show
the same quantities for a larger range of $\Im m(c)$.}
\label{gmm_sign1}
\end{center}
\vspace*{-0.3cm}
\end{figure}

The significances are shown in Figs.~\ref{gmm_sign1} for $m_H=150$ and
200~GeV, respectively, as a function of $\Im m(c)$ with $a=1$ and 
$b=0$.  As can be inferred from the
figures the maximum of the curves is slightly shifted to higher values
of $\Im m(c)/a$ compared to the corresponding
Figs.~\ref{gmm_asym1}. This is due to the increasing Higgs decay rate
with rising pseudoscalar coupling. The curves show that, even in a
best case scenario, the significance is always $\lesssim 3.5 \,
\sigma$. This asymmetry may provide only {\it evidence} for CP
violation (i.e.\ a greater than $3\, \sigma$ deviation from the SM) if
$\Im m(c) \gtrsim 1.9(0.7)$ for $m_H = 150(200)\,$GeV.

However, since one does not need to
distinguish $f_2$ and $\bar f_2$ one could also consider using jets
instead of muons, i.e. $H \to ZZ \to l^+l^-jj$, to increase the
statistics. If we use the $b \bar b$ final state, one can 
benefit from the increase by a factor $\sim 4.5$ in the branching ratio of 
the $Z$ 
boson into a $b \bar b$ pair relative to the branching ratio into a lepton 
pair. As a matter of fact a study by ATLAS~\cite{Marti:2003atl} shows that  
for a Higgs boson mass of 150 GeV with $30\,$fb$^{-1}$ it is
possible to have a Higgs signal with a significance of $2.7 \sigma$ in this
channel. So indeed one can foresee the use of this channel to add to the 
sensitivity.
\\[0.2cm]
{\bf 2. Observables which probe \boldmath{$\Re e(c)$} and/or 
\boldmath{$\Re e(b^* c)$}:} \hspace*{0.2cm}
We have constructed several observables which allow one to probe $\Re e(c)$.
For this we need an observable which is CP odd and $\tilde {\rm T}$ odd.
One possible observable is given by 
\be
O_2 = \frac{(\vec{p}_{2Z} - \vec{p}_{1Z}) \cdot 
(\vec{p}_{4H} \times \vec{p}_{3H})}{|\vec{p}_{2Z} - \vec{p}_{1Z}| 
|\vec{p}_{4H} \times \vec{p}_{3H}|} \;,
\ee
which in terms of the scattering angles reads
\br
O_2 \equiv -\sin\phi \sin\theta_1 \;.
\er
(Since $\sin \theta_1$ is always positive, one could equivalently use $\sin \phi$ as the observable and obtain the same results.)
By comparing this angular dependence with the differential angular 
decay width given in Eq.~(\ref{mmstar}), one can see that the corresponding
asymmetry should pick up the third term $\sim \eta_1 \eta_2$ of the 
contribution multiplied with $a\Re e(c)$ and the second term of the 
contribution multiplied with $\Re e(b^*c)$ and which also contains $\eta_1
\eta_2$. 
\begin{figure}[tbh]
\begin{center}
\includegraphics[width=7cm,clip=true]{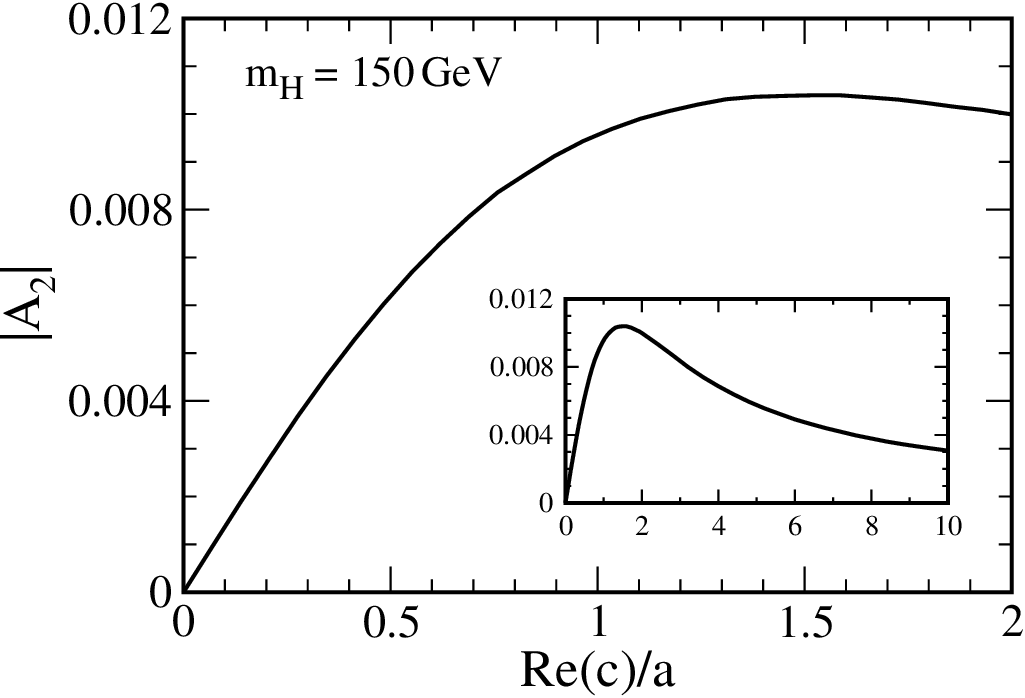} \hspace{6mm}
\includegraphics[width=7cm,clip=true]{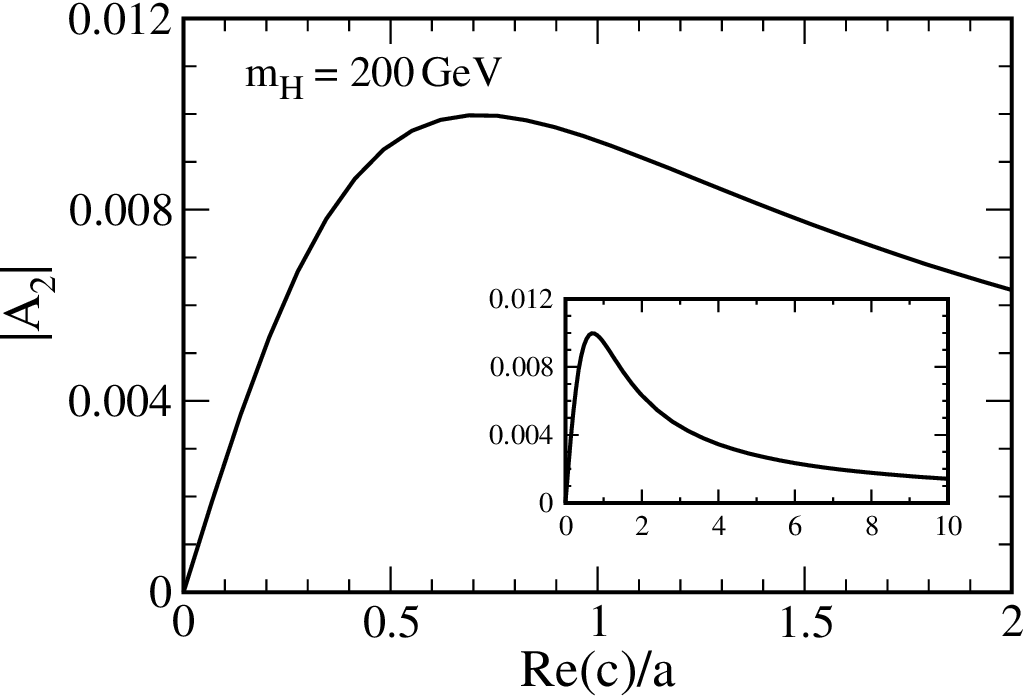}
\caption{The asymmetry ${\cal A}_2$ given by Eq.~(\ref{gmm_asym2_eq}) as a
function of the ratio $\Re e(c)/a$, for a Higgs boson of mass
$150\,$GeV ({\it left}) and 200~GeV ({\it right}). We chose $b=0$.
The inserts show the same quantities for a larger range of $\Re e(c)/a$.}
\label{gmm_asym2}
\end{center}
\vspace*{-0.3cm}
\end{figure}
And indeed we find for this asymmetry 
\br
{\cal A}_2 &=& \frac{\Gamma (O_2 > 0)-\Gamma (O_2<0)}
{\Gamma (O_2 > 0)+\Gamma (O_2<0)} \nonumber \\
&=& \frac{1}{\tilde\Gamma} \, \int \, 
d^2{\cal P} \, \left( \frac{-9\pi}{16}\right) \eta_1 \eta_2 x \gamma_b
\left[ a\Re e(c) \,\gamma_a + \Re e(b^*c) \,x
\gamma_b^2 \right] \;. 
\label{gmm_asym2_eq}
\er
By construction, for $b=0$ or to linear order in the anomalous
couplings, it is proportional to $\Re e(c)$ as expected. This
asymmetry is plotted in Figs.~\ref{gmm_asym2} as a function of $\Re
e(c)/a$ for $m_H=150$ and 200~GeV, respectively.  Since the form
factors $b,c$ are expected to be small we do not expect terms of
second order in these coefficients to have a large impact, so here and
in the following we set $b=0$. Indeed, for the asymmetry ${\cal A}_2$
with $\Re e(b^*c)\approx \Re e(c)^2 \lesssim 0.5$ the change in the
asymmetry due to neglecting $b$ is $\lesssim$ 30\%.
\begin{figure}[tbh]
\begin{center}
\includegraphics[width=7cm,clip=true]{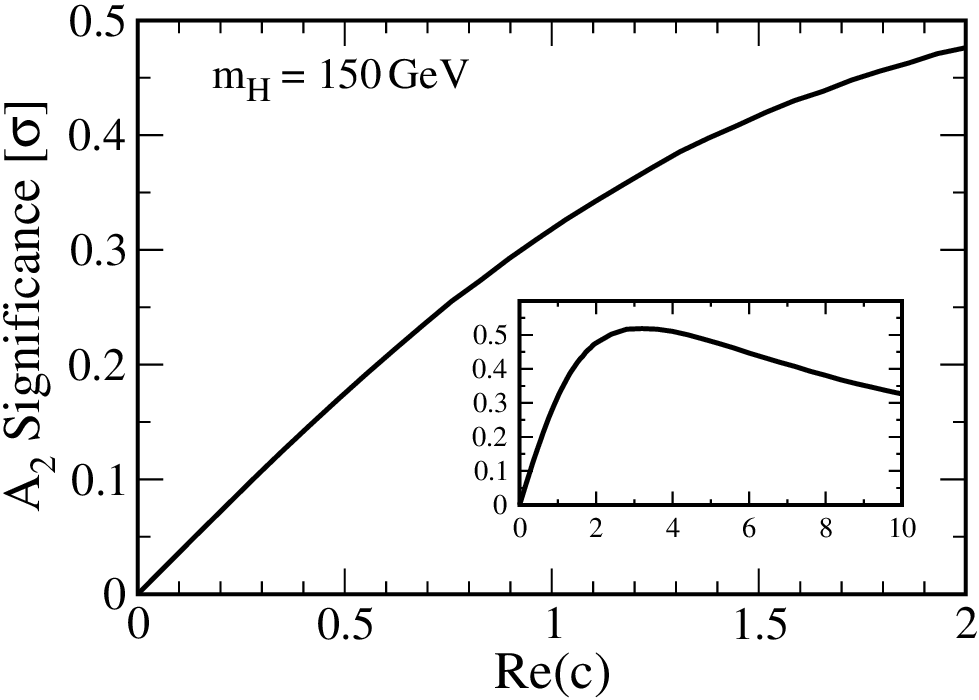} \hspace{6mm}
\includegraphics[width=7cm,clip=true]{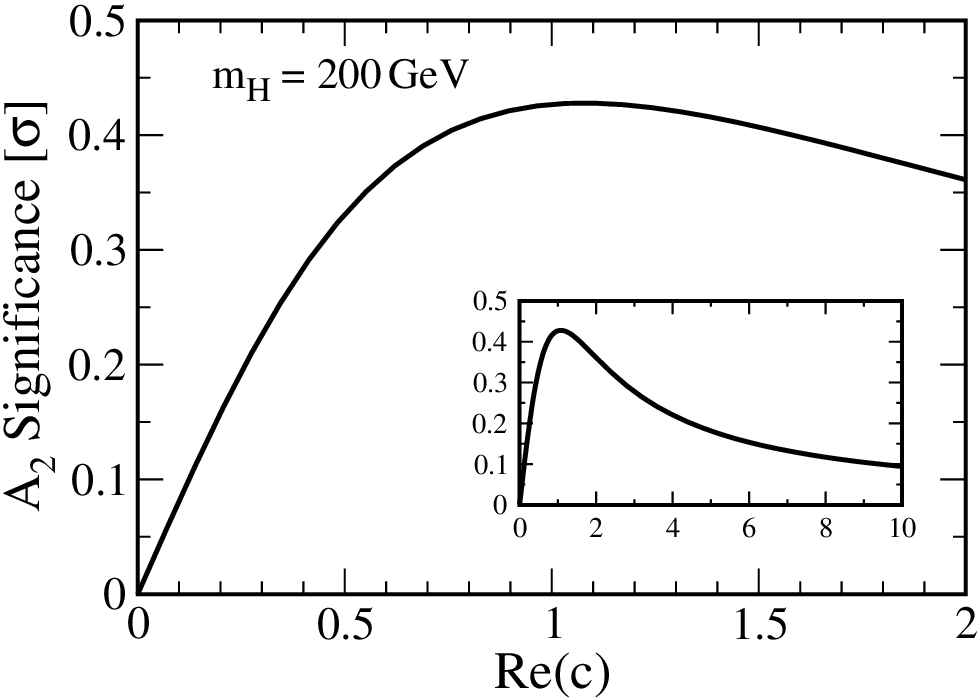}
\caption{The significances corresponding to the asymmetry ${\cal A}_2$ 
as a function of $\Re e(c)$, for a Higgs boson of mass
$150\,$GeV ({\it left}) and 200~GeV ({\it right}). We chose the other 
coupling coefficients $a=1$ and $b= 0$. The inserts show
the same quantities for a larger range of $\Re e(c)$.}
\label{gmm_sign2}
\end{center}
\vspace*{-0.3cm}
\end{figure}
Figs.~\ref{gmm_asym2} show that this asymmetry is very small, with 
values below about $\sim 0.011$, which is principally due to the 
proportionality to the small
quantity $\eta_1 \eta_2$ in Eq.~(\ref{gmm_asym2_eq}).  The
significances for the asymmetry ${\cal A}_2$ are shown in
Figs.~\ref{gmm_sign2} for the two Higgs boson mass values. With values
below about $0.55$ they are far too small to provide evidence for
CP-violation due to non-zero $\Re e(c)$. Furthermore, in this case one
cannot exploit the decay of Higgs bosons to jets since one must also
distinguish $\vec{p}_{3H}$ and $\vec{p}_{4H}$.

The smallness of the asymmetries ${\cal A}_1$ and ${\cal A}_2$ are
directly due to their proportionality to the factors $\eta_1$,
$\eta_2$. Looking at Eq.~(\ref{mmstar}), one sees that this is true for
all terms proportional to $a\, \Im m(c)$, so not much can be done to
improve on ${\cal A}_1$. However, this is not the case for terms
proportional to $a\, \Re e(c)$. So we may take our cue from the
explicit analytical expression to construct new observables for
which the asymmetry will not have these suppression factors. One such
observable is given in terms of the angles by
\br
O_3 = \cos\theta_1 \sin\theta_2 \cos\theta_2 \sin\phi \;.
\er
$O_3$ can be rewritten using the definition of $O_1$, {\it c.f.} 
Eq.~(\ref{o1}), in terms of the four three-vectors,
\br
O_3 = O_1 \, O_{3a} \, O_{3b} \;,
\er
where
\br
O_{3a} &=& \frac{(\vec{p}_{4Z} - \vec{p}_{3Z}) \cdot 
(\vec{p}_{1H} \times \vec{p}_{2H})}{|\vec{p}_{4Z} - \vec{p}_{3Z}| 
|\vec{p}_{1H} \times \vec{p}_{2H}|} \;, \nonumber \\
O_{3b} &=& \frac{(\vec{p}_{3Z} - \vec{p}_{4Z}) \cdot 
(\vec{p}_{1H} + \vec{p}_{2H})}{|\vec{p}_{3Z} - \vec{p}_{4Z}| 
|\vec{p}_{1H} + \vec{p}_{2H}|} \;.
\er
In order to exploit this observable, we have to discriminate between all 
four leptons. For the asymmetry ${\cal A}_3$, 
\br
{\cal A}_3 = \frac{\Gamma (O_3 > 0)-\Gamma (O_3<0)}
{\Gamma (O_3 > 0)+\Gamma (O_3<0)},
\er
we find analytically
\br
{\cal A}_3 &=& \frac{1}{\tilde\Gamma} \, \int \, 
d^2{\cal P} \, \left( \frac{\gamma_b x}{\pi}\right)
\left[ a \Re e(c) \, \gamma_a + \Re e(b^*c) \, x \gamma_b^2
\right]\;.
\label{gmm_asym3_eq}
\er
Note that it no longer contains the suppression factors $\eta_1$, $\eta_2$
and for  $b=0$ it probes the real part of the form factor $c$.  
By comparing the
angular structure of $O_3$ with the differential angular distribution 
Eq.~(\ref{mmstar}), one sees that the asymmetry ${\cal A}_3$ picks up the first
term in the contribution proportional to $a\Re e(c)$ and the first term 
in the one proportional to $\Re e(b^*c)$. A non-zero value
of ${\cal A}_3$ is hence an unambiguous sign of CP-violation.

%
\begin{figure}[tbh]
\begin{center}
\includegraphics[width=7cm,clip=true]{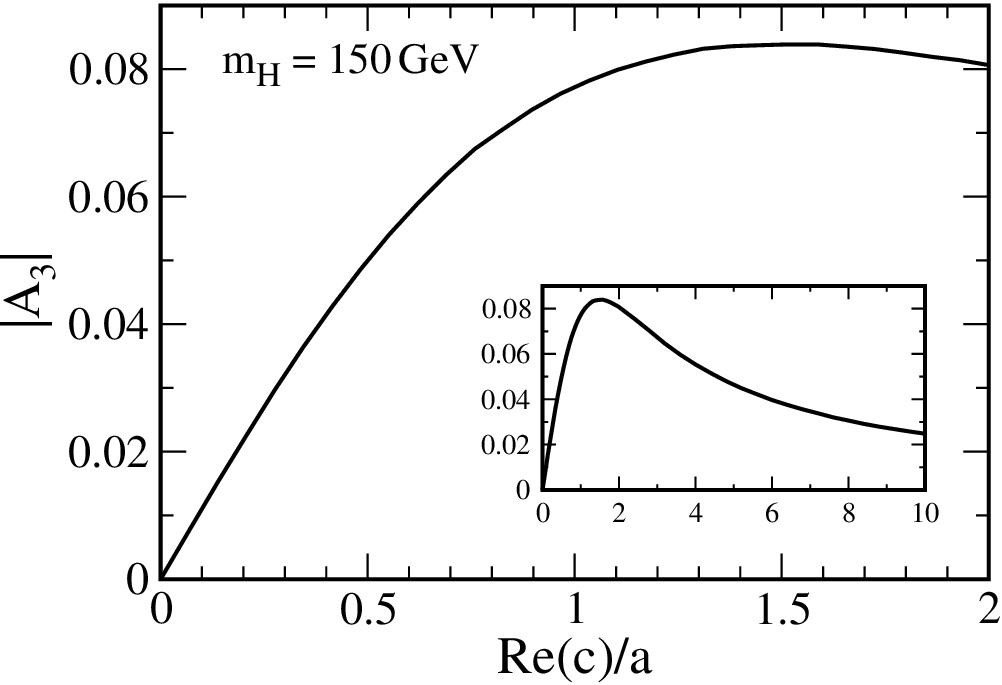} \hspace{6mm}
\includegraphics[width=7cm,clip=true]{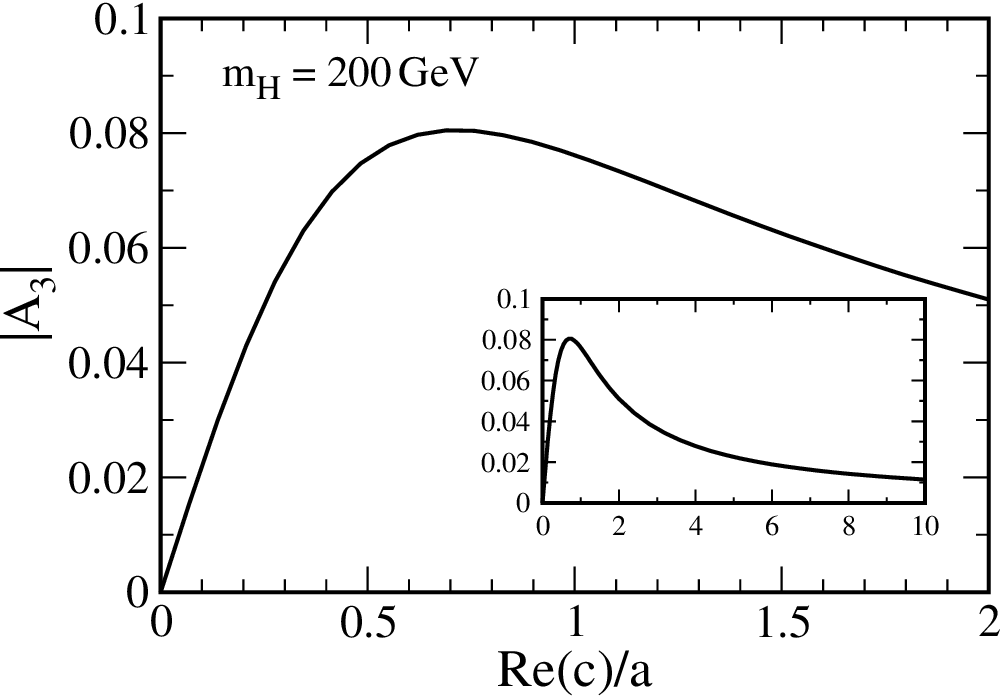}
\caption{The asymmetry ${\cal A}_3$ given by Eq.~(\ref{gmm_asym3_eq}) as a
function of the ratio $\Re e(c)/a$, for a Higgs boson of mass
$150\,$GeV ({\it left}) and 200~GeV ({\it right}). We chose $b=0$. 
The inserts show the same quantities for a larger range of $\Re e(c)/a$.}
\label{gmm_asym3}
\end{center}
\vspace*{-0.3cm}
\end{figure}
Figs.~\ref{gmm_asym3} show the asymmetry ${\cal A}_3$ for $m_H=150$ and 
$200\,$GeV, respectively, where we have taken $b=0$ for simplicity. 
With values of $\lesssim 0.09$ they are about a factor 10 larger than
those of ${\cal A}_2$. 
\begin{figure}[tbh]
\begin{center}
\includegraphics[width=7cm,clip=true]{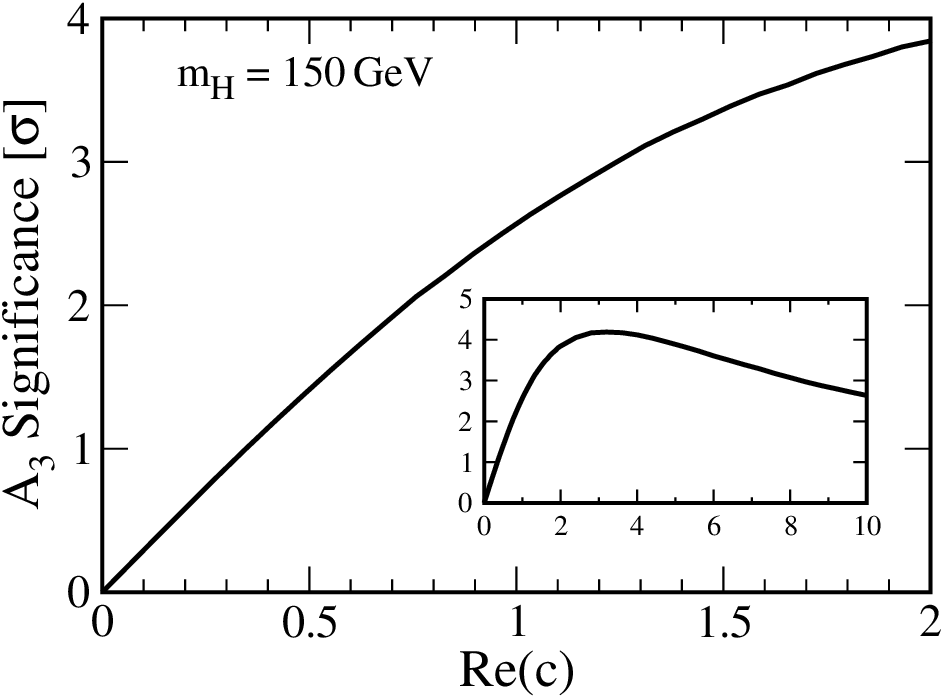} \hspace{6mm}
\includegraphics[width=7cm,clip=true]{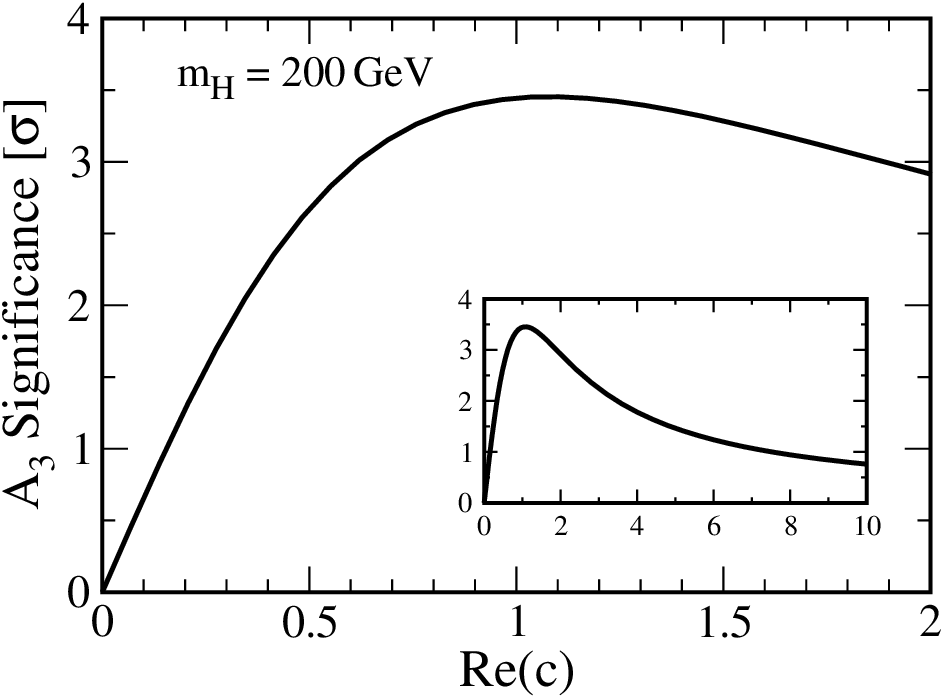}
\caption{The significances corresponding to the asymmetry ${\cal A}_3$ 
as a function of $\Re e(c)$, for a Higgs boson of mass
$150\,$GeV ({\it left}) and 200~GeV ({\it right}). We chose the other 
coupling coefficients $a=1$ and $b= 0$. The inserts show
the same quantities for a larger range of $\Re e(c)$.}
\label{gmm_sign3}
\end{center}
\vspace*{-0.3cm}
\end{figure}
The corresponding significances which should
be achievable at the LHC for this asymmetry are shown for $m_H=150$
and $200\,$GeV in Figs.~\ref{gmm_sign3}.
They are maximal at $\Re e(c) \approx 3 (1)$ for $m_H=150(200)$~GeV.
For a $150\,(200)\,$GeV Higgs boson this asymmetry would provide
evidence for CP-violation for $\Re e(c) \gtrsim 1.25\,(0.6)$ though
discovery (a $5\,\sigma$ deviation) is still out of reach. 

One should note, however, that a zero value for this asymmetry does not 
imply the absence of CP-violation, since for $b\ne 0$ it could also happen
that the contributions proportional to $a\Re e(c)$ and $\Re e(b^*c)$ cancel
and mimic CP-conservation. In order to unambiguously
show CP-violation in the $HZZ$ coupling we hence need an additional observable
to determine the two unknowns $\Re e(c)$ and $\Re e(b^* c)$. Such additional
observables are presented in the following. 
\\[0.2cm]
An observable, which probes $\Re e(c)$ alone, is given by
\br
O_4 = \frac{[(\vec{p}_{3H}\times\vec{p}_{4H})\cdot \vec{p}_{1H}]
[(\vec{p}_{3H}\times\vec{p}_{4H})\cdot (\vec{p}_{1H}\times\vec{p}_{2H})]}{|\vec{p}_{3H}+\vec{p}_{4H}|^2|\vec{p}_{1H}+\vec{p}_{2H}||\vec{p}_{3Z}-\vec{p}_{4Z}|^2|\vec{p}_{1Z}-\vec{p}_{2Z}|^2/16} \;.
\er
In terms of the angles it reads
\br
O_4 = \sin^2\theta_1 \sin^2\theta_2 \sin\phi \cos\phi \;.
\er
(Again, since $\sin^2 \theta_{1,2}$ are always positive, this is
equivalent to using an observable $\sin 2 \phi$.)  This coupling
structure appears in the decay width only in the contribution which is
proportional to $a\Re e(c)$, {\it c.f.} Eq.~(\ref{mmstar}), so that we
can expect the corresponding asymmetry to probe CP-violation due to
simultaneous non-vanishing form factors $a$ and $c$ unambiguously.
Indeed the asymmetry is given by
\br
{\cal A}_4 &=& \frac{\Gamma (O_4 > 0)-\Gamma (O_4<0)}
{\Gamma (O_4 > 0)+\Gamma (O_4<0)} \nonumber\\
&=& \frac{1}{\tilde\Gamma} \, \int \, 
d^2{\cal P} \, \left( \frac{-2}{\pi}\right) a \Re e(c) x\,\gamma_b
\;.
\label{gmm_asym4_eq}
\er

\begin{figure}[htb]
\begin{center}
\includegraphics[width=7cm,clip=true]{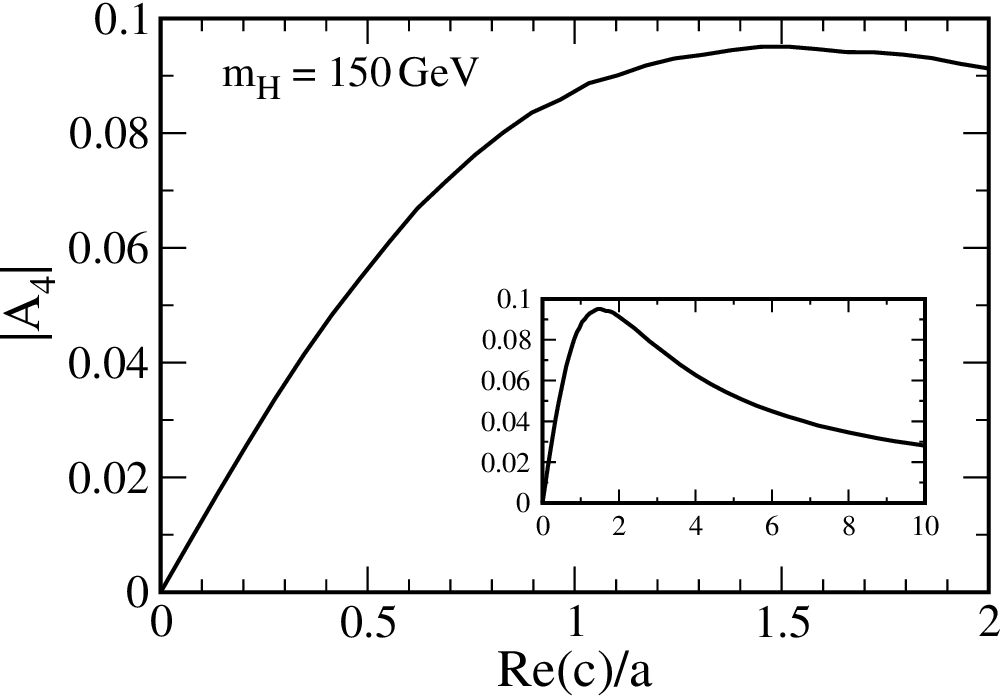} \hspace{6mm}
\includegraphics[width=7cm,clip=true]{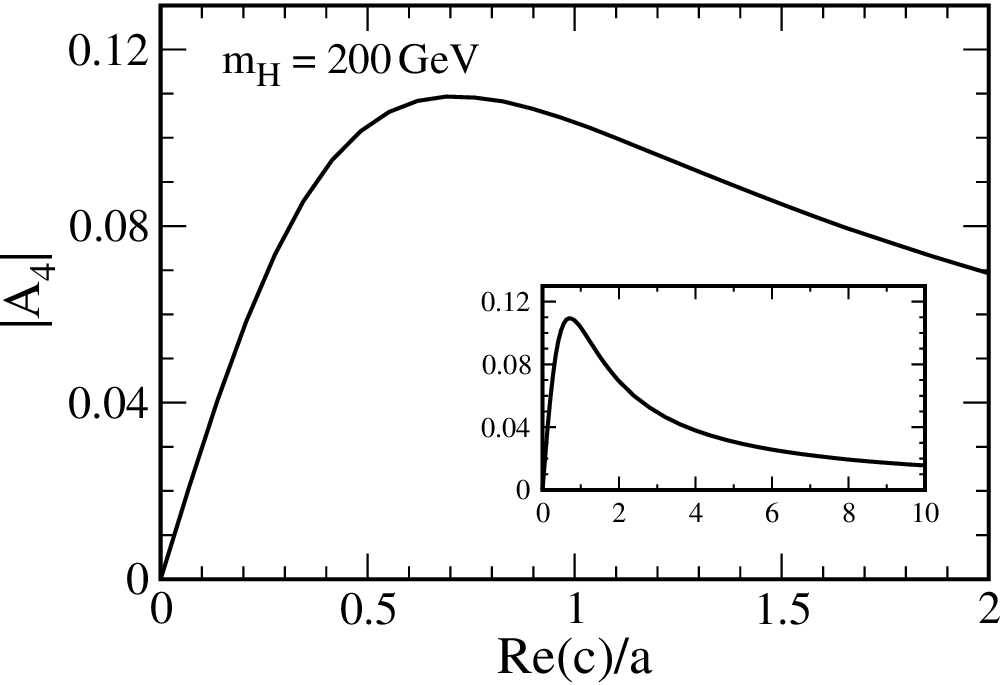}
\vspace*{-0.1cm}
\caption{The asymmetry ${\cal A}_4$ given by Eq.~(\ref{gmm_asym4_eq}) as a
function of the ratio $\Re e(c)/a$, for a Higgs boson of mass
$150\,$GeV ({\it left}) and 200~GeV ({\it right}). 
The inserts show the same quantities for a larger range of $\Re e(c)/a$.}
\label{gmm_asym4}
\end{center}
\vspace*{-0.2cm}
\end{figure}
\begin{figure}[htb]
\begin{center}
\includegraphics[width=7cm,clip=true]{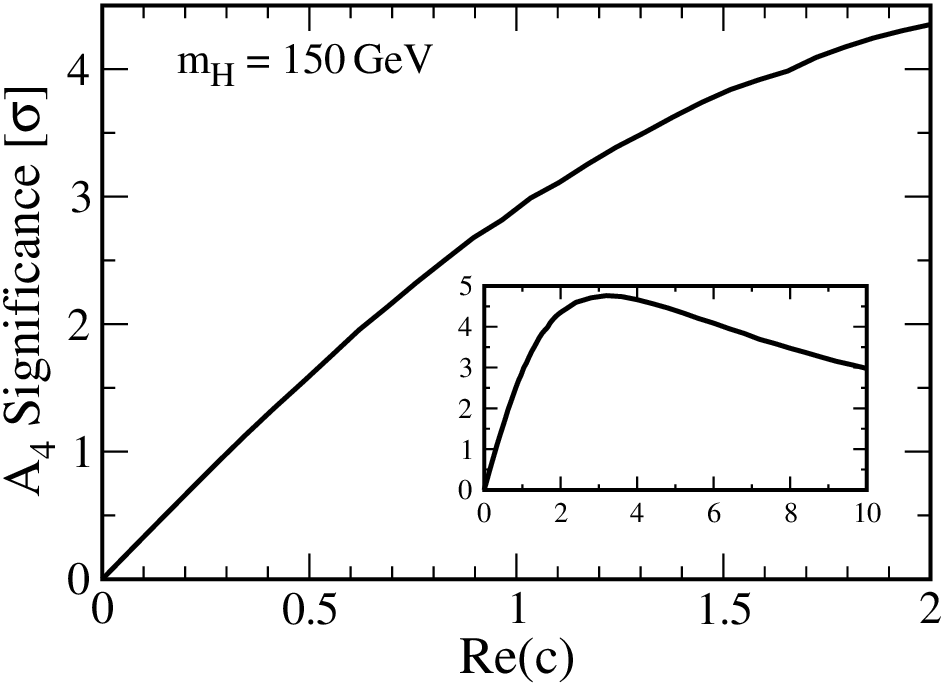} \hspace{6mm}
\includegraphics[width=7cm,clip=true]{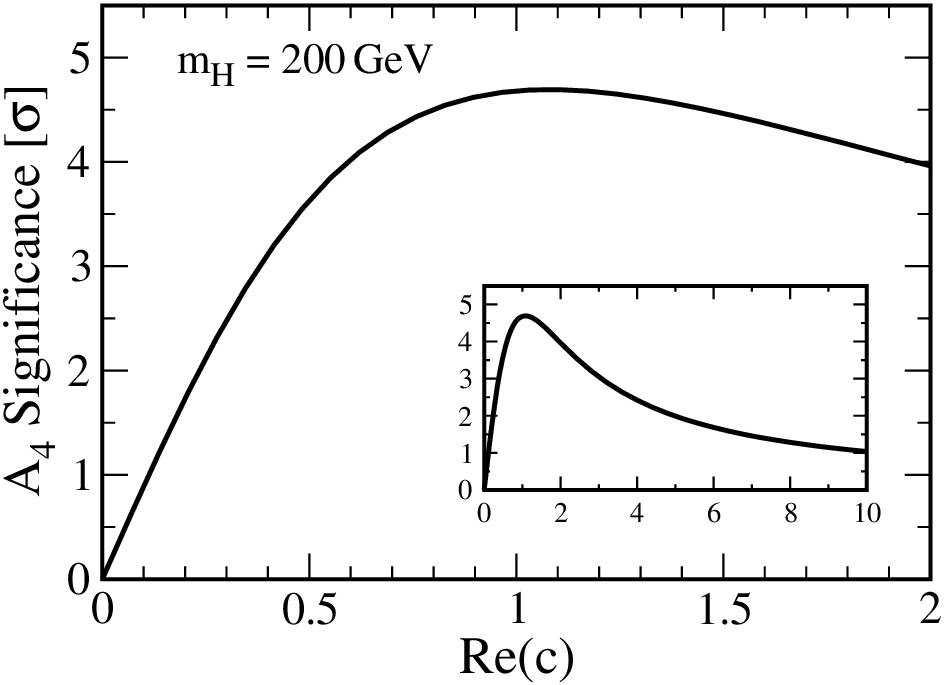}
\vspace*{-0.1cm}
\caption{The significances corresponding to the asymmetry ${\cal A}_4$ 
as a function of $\Re e(c)$, for a Higgs boson of mass
$150\,$GeV ({\it left}) and 200~GeV ({\it right}). We chose the other 
coupling coefficients $a=1$ and $b= 0$. The inserts show
the same quantities for a larger range of $\Re e(c)$.}
\label{gmm_sign4}
\end{center}
\vspace*{-0.4cm}
\end{figure}
Furthermore, as can be inferred from Figs.~\ref{gmm_asym4}, which show
the asymmetry for $m_H=150$ and 200~GeV as a function of $\Re e(c)/a$, 
the asymmetries are larger than those of ${\cal A}_2$ and ${\cal A}_3$,
with maximal values of up to $\sim 0.11$. 
The significances which may be achieved at the LHC are shown in
Figs.~\ref{gmm_sign4}. They reach values of up to almost $5$ for
$m_H=150,200\,$GeV so that this observable may probe CP-violation in an 
unambiguous way at the LHC for sufficiently large values of $\Re e(c)$. 
As can be inferred from Figs.~\ref{gmm_sign4}, for a $150\,$GeV Higgs boson
evidence for a non-zero $\Re e(c)$ is possible for $\Re e(c)\gtrsim
1$, while for a $200\,$GeV Higgs boson evidence is already possible
for $\Re e(c)\gtrsim 0.4$.

\begin{figure}[tbh]
\begin{center}
\includegraphics[width=7cm,clip=true]{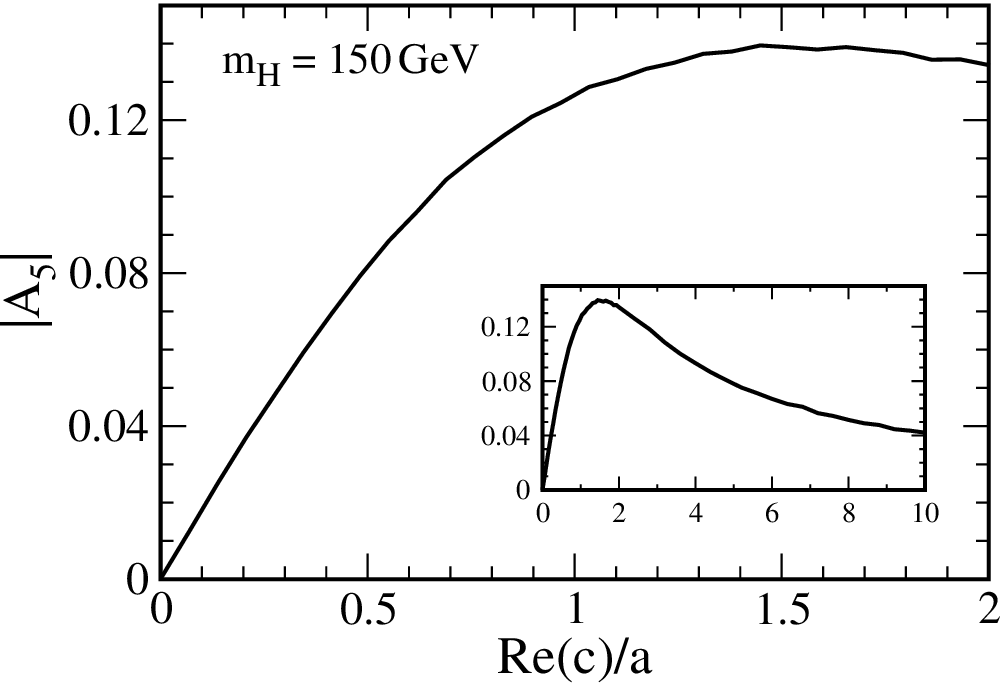} \hspace{6mm}
\includegraphics[width=7cm,clip=true]{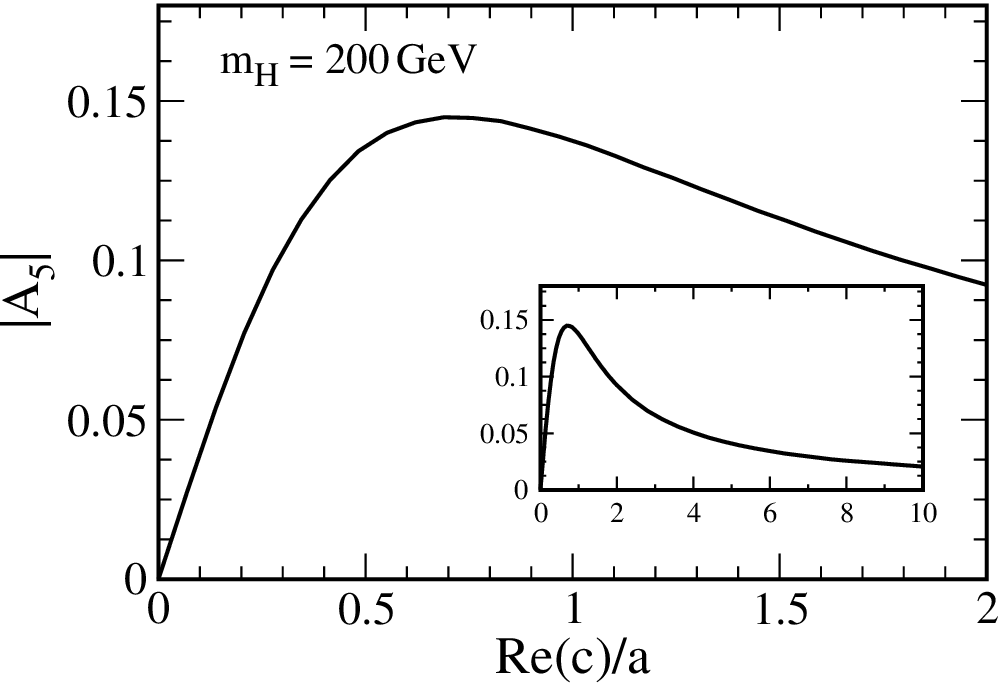}
\caption{The asymmetry ${\cal A}_5$ given by Eq.~(\ref{gmm_asym5_eq}) as a
function of the ratio $\Re e(c)/a$, for a Higgs boson of mass
$150\,$GeV ({\it left}) and 200~GeV ({\it right}). We chose $b=0$. 
The inserts show the same quantities for a larger range of $\Re e(c)/a$.}
\label{gmm_asym5}
\end{center}
\vspace*{-0.3cm}
\end{figure}

\begin{figure}[tbh]
\begin{center}
\includegraphics[width=7cm,clip=true]{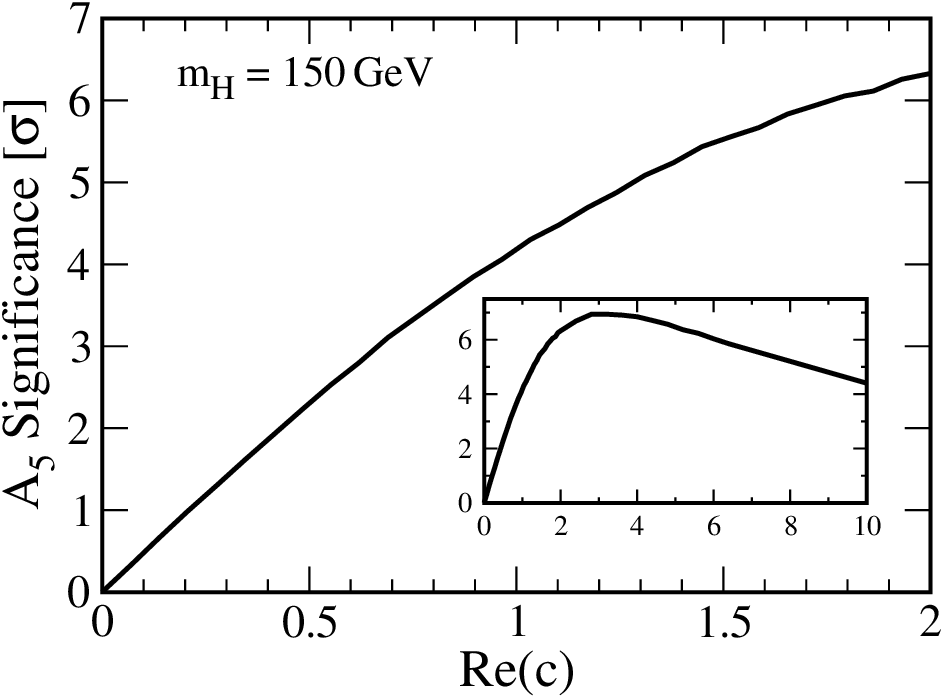} \hspace{6mm}
\includegraphics[width=7cm,clip=true]{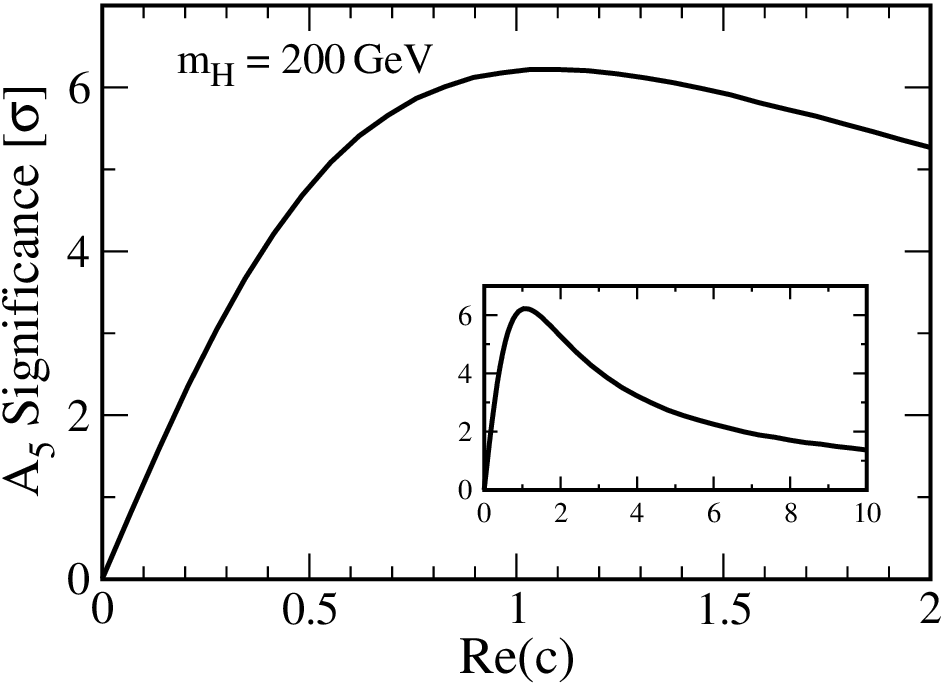}
\caption{The significances corresponding to the asymmetry ${\cal A}_5$ 
as a function of $\Re e(c)$, for a Higgs boson of mass
$150\,$GeV ({\it left}) and 200~GeV ({\it right}). We chose the other 
coupling coefficients $a=1$ and $b= 0$. The inserts show
the same quantities for a larger range of $\Re e(c)$.}
\label{gmm_sign5}
\end{center}
\vspace*{-0.3cm}
\end{figure}

Alternatively one may use a combination of $O_3$ and $O_4$ to test
CP-violation due to non-vanishing $a\Re e(c)$ and/or $\Re
e(b^*c)$. For example, in terms of the angles a possible observable
$O_5$ is given by 
\br 
O_5 = \sin\theta_1 \sin\theta_2 \sin\phi
[\sin\theta_1 \sin\theta_2\cos\phi - \cos\theta_1 \cos\theta_2 ] 
\er
and can be constructed from the three-vectors by \br O_5 =
\frac{[(\vec{p}_{4H}\times\vec{p}_{3H})\cdot\vec{p}_{1H}][(\vec{p}_{1Z}
-\vec{p}_{2Z})\cdot\vec{p}_{3Z}]}{|\vec{p}_{3H}+\vec{p}_{4H}||\vec{p}_{3Z}-\vec{p}_{4Z}|^2
|\vec{p}_{1Z}-\vec{p}_{2Z}|^2/8} \;.  \er The related asymmetry \br
{\cal A}_5 = \frac{\Gamma (O_5 > 0)-\Gamma (O_5<0)} {\Gamma (O_5 >
0)+\Gamma (O_5<0)}
\label{gmm_asym5_eq}
\er
is shown in Figs.~\ref{gmm_asym5} for $m_H=150,200$~GeV and yields the 
largest values among the asymmetries discussed so far, up to $\sim 0.15$. 
In Figs.\ref{gmm_sign5} we show the corresponding
significances. We see that for a $150\,(200)\,$GeV Higgs boson, this
asymmetry would provide evidence for CP-violation for $\Re e(c)
\gtrsim 0.66\,(0.25)$ and discovery of CP-violation for $\Re e(c)
\gtrsim 1.28(0.52)$.
%
\\
{\bf 3. An observable which probes \boldmath{$\Im m(b)$}:} \hspace*{0.2cm}
For completeness, we also present an observable that probes the imaginary part 
of the CP-even form factor $b$. It is given by the following combination of 
three-vectors
\br
O_6 &=& \frac{[(\vec{p}_{1Z}-\vec{p}_{2Z})\cdot (\vec{p}_{3H}+\vec{p}_{4H})]
[(\vec{p}_{3H}\times\vec{p}_{4H})\cdot \vec{p}_{1H}]}{|\vec{p}_{1Z}-\vec{p}_{2Z}|^2 |\vec{p}_{3H}+\vec{p}_{4H}|^2 |\vec{p}_{3Z}-\vec{p}_{4Z}|/4}
\nonumber\\
&=& \sin\theta_1 \cos\theta_1 \sin\theta_2 \sin\phi \;.
\er

\begin{figure}[tbh]
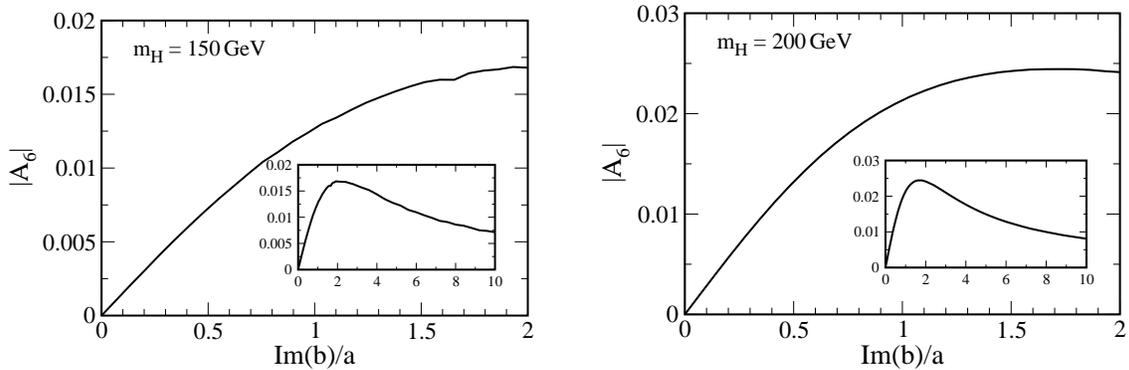

\begin{center}
\includegraphics[width=7cm,clip=true]{a6_mag_150.eps} \hspace{6mm}
\includegraphics[width=7cm,clip=true]{a6_mag_200.eps}
\caption{The asymmetry ${\cal A}_6$ given by Eq.~(\ref{gmm_asym6_eq}) as a
function of the ratio $\Im m(b)/a$, for a Higgs boson of mass
$150\,$GeV ({\it left}) and 200~GeV ({\it right}). We chose $c=0$. 
The inserts show the same quantities for a larger range of $\Im m(b)/a$.}
\label{gmm_asym6}
\end{center}
\vspace*{-0.3cm}
\end{figure}

\begin{figure}[tbh]
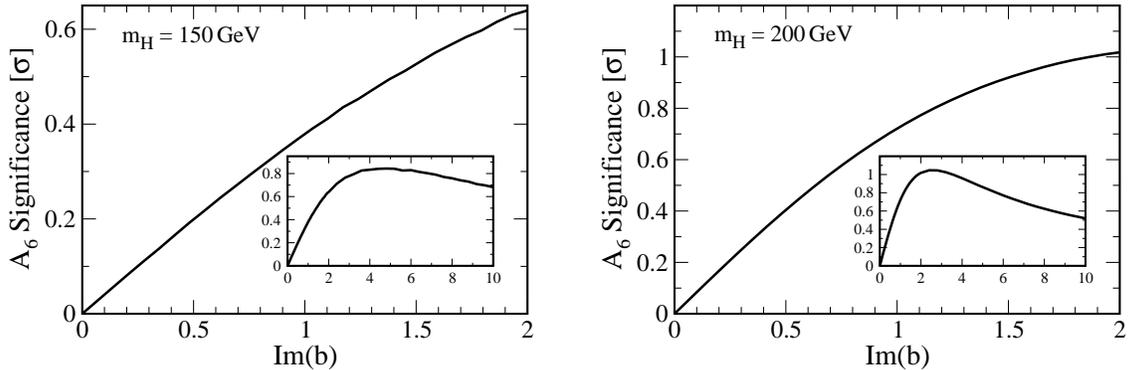

\begin{center}
\includegraphics[width=7cm,clip=true]{a6_sig_150.eps} \hspace{6mm}
\includegraphics[width=7cm,clip=true]{a6_sig_200.eps}
\caption{The significances corresponding to the asymmetry ${\cal A}_6$ 
as a function of $\Im m(b)$, for a Higgs boson of mass
$150\,$GeV ({\it left}) and 200~GeV ({\it right}). We chose the other 
coupling coefficients $a=1$ and $c= 0$. The inserts show
the same quantities for a larger range of $\Im m(b)$.}
\label{gmm_sign6}
\end{center}
\vspace*{-0.3cm}
\end{figure}

\noindent
And the asymmetry reads analytically
\br
{\cal A}_6 &=& 
 \frac{\Gamma (O_6 > 0)-\Gamma (O_6<0)}
{\Gamma (O_6 > 0)+\Gamma (O_6<0)} \nonumber\\
&=& \frac{1}{\tilde\Gamma} \, \int \, 
d^2{\cal P} \,\frac{3}{8} \eta_2 \, a \Im m(b) \, x\gamma_b^2 \;.
\label{gmm_asym6_eq}
\er
Figs.~\ref{gmm_asym6} and \ref{gmm_sign6} show the corresponding asymmetries
and significances. 
Notice that once again, the asymmetry is proportional to the small
factor $\eta_2$ and is therefore rather small, {\it i.e.} $\lesssim
0.025$. Correspondingly this observable does not provide a good
significance (only reaching values of about $1$), so that the
extraction of $\Im m(b)$ from this observable does not seem to be
feasible at the LHC. Unfortunately, since this small factor is present
in all the relevant terms in Eq.~(\ref{mmstar}), all asymmetries that
one can construct to probe this coefficient will be similarly small. 
%

Refs.\cite{Skjold:1993jd,Skjold:1994qn} also consider
reweighting observables with the product of the energy differences
between the paired leptons, i.e.\ $(E_2-E_1)(E_4-E_3)$. In our
notation, this product can be written,
\br
(E_2-E_1)(E_4-E_3) = \gamma_1 \gamma_2 \beta_1 \beta_2 m_1 m_2 \cos \theta_1 \cos \theta_2 \;.
\er
So this procedure places more importance on events with highly boosted
$Z$ bosons and/or events where the lepton is emitted along the line of
the parent $Z$ boson's direction of travel. Only the latter would
affect our asymmetries and is similar (in principle and effect) to
making a different choice of the observable, $O_i$. Thus, the procedure 
adopted by these authors is analogous to what we have done. 
\\[0.2cm]
\indent
In summary, using the observables $O_1$, $O_3$, $O_4$ and $O_5$ and
their corresponding asymmetries at the LHC, we can in principle place
limits on (or provide evidence for) CP-violation due to the
simultaneous presence of CP-even and CP-odd form factors. For both a
$150\,$GeV and $200\,$GeV Higgs boson, $O_1$, $O_3$, $O_4$ and $O_5$
all provide evidence (and $O_5$ potential discovery) for some values
of the additional non-SM couplings. Unfortunately, 
$O_2$ and $O_6$ are rather
insensitive (due to the requirement of vector-axial interference) and
cannot be used to place useful limits on additional couplings. The
observables $O_3$ and $O_5$ can not unambiguously rule out
CP-violation, since their dependence also on $\Re e(b^*c)$ may provide
an accidental cancellation with the terms proportional to $a \Re e(c)$. 
However, $O_4$ only depends on $a \Re
e(c)$ and can thus test violation of the CP quantum numbers due to
non-vanishing $a$ and $c$. With the three observables $O_{3,4,5}$ at
hand we can furthermore also extract the value of $\Re e(b^*c)$ and
finally do consistency tests as well as reduce the effect of
experimental errors.

From a theoretical perspective, these asymmetries (if measurable) are
sufficient to determine all the form factors $a$, $b$ and $c$ of our
general CP-violating $HZZ$ coupling, with real and imaginary parts.
This is summarized in Table~\ref{obsdep}, which shows the various
dependencies of the described observables on the form factors. We have
6 observables for the five unknowns $a$, $\Re e(b)$, $\Im m(b)$, $\Re
e(c)$ and $\Im m(c)$.  If we furthermore assume that any product of
$b$ and $c$ is very small (if $b$ and $c$ are loop suppressed or
suppressed by some scale of new physics) we may neglect their
simultaneous influence in the asymmetries ${\cal A}_2$, ${\cal A}_3$
and ${\cal A}_5$ and only require two of the asymmetries ${\cal A}_3$
to ${\cal A}_5$ (${\cal A}_2$ not being of much use due to its
smallness) to extract $a$ and $\Re e(c)$, while relying on ${\cal
A}_1$ and ${\cal A}_6$ for $\Im m(c)$ and $\Im m(b)$
respectively. However, the analysis done here and the smallness of the
asymmetries implies that only $\Re e(c)$, and possibly $\Im m(c)$, are
likely to be seen if non-zero. Also note that many of these
asymmetries are highly correlated with one another.

\begin{table}
\begin{center}
\begin{tabular}{|c||c|c|c|c|c|}
\hline
Asymmetry/form factor & $a$ & $\Re e(b)$ & $\Im m(b)$ & $\Re e(c)$ & $\Im m(c)$
\\ \hline \hline
${\cal A}_1$ & x & & & & x\\ \hline
${\cal A}_2$ & x & (x) & (x) & x & (x) \\ \hline
${\cal A}_3$ & x & (x) & (x) & x & (x) \\ \hline
${\cal A}_4$ & x & & & x & \\ \hline
${\cal A}_5$ & x & (x) & (x) & x & (x)\\ \hline
${\cal A}_6$ & x & & x & & \\ \hline
\end{tabular}
\end{center}
\caption{The dependence of the asymmetries ${\cal A}_1$
to ${\cal A}_6$ on the form factors $a,b,c$ of the general $HZZ$
coupling Eq.~(\ref{param}). $({\rm x})$ denotes a dependence which is
suppressed if the additional form factors are small.}
\label{obsdep}
\end{table}
%
Of course, the final feasibility of detecting or placing limits on
non-SM form factors depends on the real experimental
environment. We simulated this here by taking the values given by the
ATLAS studies. Any further refinement is beyond the scope of this
study, but we have shown here the utility of these observables in
providing unambiguous information on possible non-SM terms in the
$HZZ$ coupling and the consequent CP-violation. Any evidence or
discovery of CP-violation crucially depends on the size of the
non-SM form factors. Irrespective of this one may use
these observables to place experimental limits on their values. \\[0.2cm]
\section{Kinematical distributions as a probe of CP-violation}
\label{sectfive}
As we have seen, the asymmetries discussed in section~\ref{sectfour}
are most useful for a Higgs boson with mass $m_H \gsim 2 m_Z$. For a
lighter Higgs boson the rates are much smaller and the significance
may not be sufficient for identifying CP-violation or setting
satisfactory limits. In this case, one must rely on fitting shapes of
kinematic distributions that depend on the CP character of the Higgs 
boson. From the discussions in the literature it is clear that the
angle $\phi$ between the planes of the two fermion pairs coming from
the decays of the $Z$ bosons, and the polar angle of the fermions
$f_1$ or $f_2$ in the rest frames of the $Z$ bosons, $\theta_i$ ($i=1,2$), 
are suitable variables \cite{Choi:2002jk} (see Fig.~\ref{gmmm_angles}).
\\[0.2cm]
%
{\underline{\bf 1) The angular distribution in \boldmath{$\phi\,$}:}} 
\hspace*{0.2cm}
In the decay process Eq.~(\ref{hdecayzz}), let us consider the azimuthal 
angular distribution 
$d\Gamma/d\phi$. Integrating Eq.~(\ref{mmstar}) over $\theta_1, \theta_2$
and taking a CP-violating coupling with $a$ and $c$ 
non-zero\footnote{The expression
with all three coupling coefficients $a$, $b$ and $c$ non-zero is given in the
Appendix.} we find 
\br
\frac{d\Gamma}{d\phi} &\sim& b_1 + b_2\cos\phi + b_3 \sin\phi + b_4 \cos 2\phi
+ b_5 \sin 2\phi \;, 
\er
where $b_i$ ($i=1,...,5$) are functions of $m_H$ and $m_Z$ in terms of 
$\gamma_a,\gamma_b$,
\br
b_1 &=& a^2 (2+\gamma_a^2) + 8 |c|^2 x^2 \gamma_b^2
\nonumber \\
b_2 &=& -\frac{9\pi^2}{32} \, a^2 \, \eta_1\eta_2 \gamma_a \nonumber\\
b_3 &=& \frac{9\pi^2}{16}\, a \Re e(c) \, \eta_1\eta_2 x \gamma_a \gamma_b
\\
b_4 &=& \frac{a^2}{2} -2 |c|^2 x^2 \gamma_b^2 \nonumber\\
b_5 &=&  -2 a \Re e(c)\, x \gamma_b\;. 
\nonumber
\label{phicoeff}
\er
Whereas the purely SM case ($a=1,b=c=0$) shows a distribution (see also
Ref.~\cite{Barger:1993wt})
\br
\frac{d\Gamma}{d\phi} &\sim& 1 + a_2 \cos\phi + a_4 \cos 2\phi\;, \nonumber\\
a_2 &=& -\frac{9\pi^2}{32} \eta_1 \eta_2 \frac{\gamma_a}{2+\gamma_a^2} \\
a_4 &=& \frac{1}{2}\frac{1}{2+\gamma_a^2} \;, \nonumber
\er
in the purely pseudoscalar case ($a=b=0$, $c\ne0$) we have
\br
\frac{d\Gamma}{d\phi} \sim 1 - \frac{1}{4} \cos 2\phi \;.
\er
In the CP violating case the inclusion of contributions from both the
scalar and pseudoscalar couplings alters the angular behaviour via the
occurrence of $\sin \phi$ and $\sin 2 \phi$ terms, and a reweighting of
the other terms. Knowing the Higgs mass from previous measurements,
any deviation from the predicted distribution in the purely
scalar/pseudoscalar case will be indicative of CP violation.
%
\begin{figure}[tbh]
\begin{center}
\includegraphics[width=8cm,clip=true]{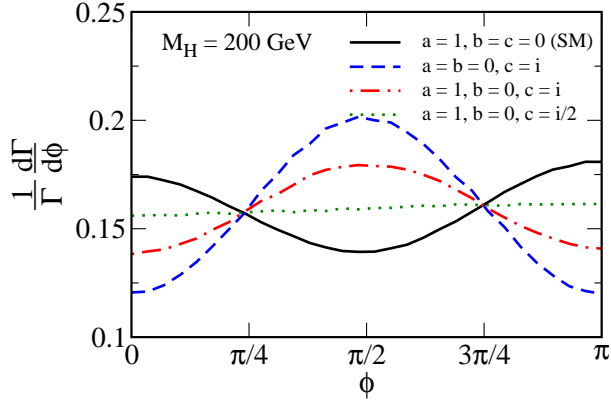}
\caption{The normalized differential width for \mbox{$H \rightarrow
Z^{(*)} Z \rightarrow (f_1\bar{f}_1) (f_2\bar{f}_2)$} with respect to
the azimuthal angle $\phi$. The solid (black) curve shows the
\sm case ($a=1$, $b=c=0$) while the dashed (blue) curve is a
pure CP-odd state ($a=b=0$, $c=i$). The dot-dashed (red) curve and the
dotted (green) curve are for states with CP violating couplings $a=1$,
$b=0$ with $c=i$ and $c=i/2$, respectively.}
\label{gmm_phi}
\end{center}
\vspace*{-0.4cm}
\end{figure}
This can be inferred from Fig.~\ref{gmm_phi} which shows the azimuthal
angular distribution for $m_H=200\,$GeV in the SM case, for a purely
CP-odd Higgs boson and for two CP violating cases. The purely CP-odd
curve will always show the same behaviour independently of the value
of $c$ since the curves are normalized to unit area. Therefore a
special value of $c$ could not fake the flattening of the curve
appearing in the CP violating examples. This flattening even leads to
an almost constant distribution in $\phi$ for the case $c/a=i/2$. It
should be kept in mind though, that this method cannot be applied for
very large Higgs masses where the $\phi$ dependence is washed out. One
must also beware of degenerate Higgs bosons of opposite CP; since the
decay products are the same, they will both contribute to the rate and
must be summed coherently, possibly mimicking the effect seen above.

\begin{figure}[tbh]
\begin{center}
\includegraphics[width=8cm,clip=true]{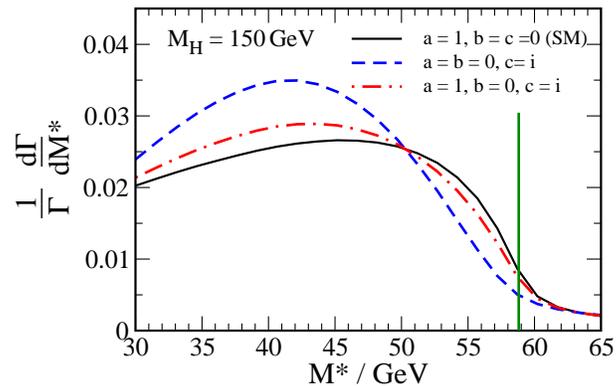}
\caption{The normalized differential width for \mbox{$H \rightarrow
Z^{(*)} Z \rightarrow (f_1\bar{f}_1) (f_2\bar{f}_2)$} with respect to
the virtuality of the (most) off-shell Z-boson $M^*$. The solid
(black) curve shows the \sm case ($a=1$, $b=c=0$) while the dashed
(blue) curve is a pure CP-odd state ($a=b=0$, $c=i$). The dot-dashed
(red) curve is for states with CP violating couplings $a=1$, $b=0$
with $c=i$. The vertical green line represents the nominal threshold
at $m_H-m_Z$.}
\label{gmm_thresh}
\end{center}
\end{figure}

This procedure is similar to that of Refs.~\cite{Buszello:2002uu,Bluj} where
log-likelihood functions were constructed and minimised to extract the
coefficients in the vertex or yield exclusion contours.\\[0.2cm]
%
{\bf \underline{2) The angular distribution in \boldmath{$\theta_i\,$}:}} 
\hspace*{0.2cm}
Integrating Eq.~(\ref{param}) over $\phi$ and $\cos\theta_2$ provides
a distribution in $\cos\theta_1$. For the CP violating case $a,c\ne 0,
\, b=0$ we find,
\br
\hspace*{-0.4cm}
\frac{d\Gamma}{d\cos\theta_1} &\sim& a^2 \left[(\gamma_a^2-1) 
\sin^2\theta_1 + 2 \right] + 
4 |c|^2 x^2 \gamma_b^2 (1+\cos^2\theta_1)
 - 8 a \Im m(c) \eta_1 x \gamma_b\cos\theta_1 . 
\er
In the purely SM case we recover,
\br
\frac{d\Gamma}{d\cos\theta_1} &\sim& \sin^2\theta_1 + \frac{2}{\gamma_a^2-1},
\er
which for large Higgs boson masses ($\gamma_a \to \infty$) reproduces the 
well-known behaviour $\sim \sin^2\theta_1$. In contrast, in the purely CP 
odd case we have
\br
\frac{d\Gamma}{d\cos\theta_1} \sim 1 + \cos^2 \theta_1 \;.
\er
CP violation is manifest by a linear dependence on
$\cos\theta_1$. However, due to the proportionality to $\eta_1$ the CP
violating effect in the angular distribution is small, which
is reflected also in the smallness of the asymmetry ${\cal A}_1$. See
also the discussion in Section~\ref{sectfour} and
Fig.~\ref{gmm_theta}.\\[0.2cm]
%
{\bf \underline{2) The threshold distribution:}} 
\hspace*{0.2cm}
In principle, information about the form factors of the $HZZ$ vertex
is also encoded in the dependence of its partial width on the
virtuality of the $Z$-bosons~\cite{Choi:2002jk}. In particular,
looking at Eq.~(\ref{htoz1z2}) one sees that only the term
proportional to $a^2$ contains a linear dependence in $\beta$. This is due
to there being no momentum dependence in the SM $HZZ$ vertex, in
contrast to the additional non-SM terms of Eq.~(\ref{param}); the 
single $\beta$ arises from the phase space. Consequently, one can distinguish
a CP-even Higgs boson from a CP-odd Higgs boson decaying to $ZZ^*$ by
examining the threshold behaviour since the CP-even excitation curve will be
much steeper. This is illustrated in Fig.~\ref{gmm_thresh} where one
can clearly see the steeper dependence on the virtuality $M^*$ of the
most off-shell Z-boson for the CP-even case compared to the CP-odd case.

However, this behaviour near threshold will be dominated by whichever
term has the lowest power of $\beta$. So when one has a Higgs boson of
mixed CP, the SM term will always dominate at threshold. This is also
shown in Fig.~\ref{gmm_thresh} where the curve for the CP-violating
case sits almost on top of the SM curve near threshold. So while the
threshold dependence is very good at distinguishing a pure CP-even
Higgs boson from a pure CP-odd one, it is unfortunately not very
helpful for distinguishing a CP-violating Higgs from the SM case.

\section{Conclusions}
\label{sectconcl}
In this work we have studied the process $H\to ZZ^{(*)}\to 4l$,
($l=e,\mu$) at the LHC to determine how well a general CP
violating $HZZ$ coupling can be tested. 

We examined the dependence of the partial width on non-SM form
factors. By making use of the expected numbers of SM signal and
background events, after cuts, provided by the ATLAS experiment, we produced
exclusion plots for these non-SM form factors. We
demonstrated that while large non-SM form factors may cause large
deviations, it is difficult to distinguish their effect from an
enhanced (or diminished) SM coupling.

We then presented asymmetries which are non-vanishing when non-SM form
factors are present in the $HZZ$ coupling. We found a set of observables 
which, in principle, allows the extraction of the real and imaginary parts 
of all the complex form factors in the non-SM part of the $HZZ$ vertex, if the 
significances are large enough.  We analysed
these asymmetries in the context of the ATLAS $H\to ZZ^{(*)}\to
4l$ study, and found that some of these asymmetries may be large
enough to provide evidence of CP violation and in some cases even
discovery, depending of course on the specific values of the CP
violating contributions. In any case, these asymmetries will be useful
in putting limits on any possible extra $HZZ$ couplings beyond the
tree-level SM, and deserve further experimental analysis.

Furthermore, we presented an analytic formula for the partial width
with full dependence on the final state azimuthal and polar angles,
and demonstrated that the angular distributions may be
exploited for Higgs boson masses below the threshold. Indeed, the
azimuthal angle between the two decay planes of the $Z$ bosons is
sensitive to CP violation if the Higgs boson mass is not too large.

\section{Acknowledgments}
We wish to acknowledge with thanks useful discussions with D.~Choudhury, 
A.~Nikitenko, P.~Osland, M.~Schumacher, A.~Str\"assner and D.~Zeppenfeld.  
R.M.G.\ and M.M.\ wish to acknowledge support from the Indo
French Centre for Promotion of Advanced Research Project 3004-2. We
also thank the funding agency Board for Research in Nuclear Sciences
and the organizers of the 9th Workshop on High Energy Physics
Phenomenology (WHEPP9), held in Bhubaneswar where part of this work
was discussed. We are grateful to M.~Spira for the careful reading of
the manuscript.

\section{Appendix}
For the process $H\to ZZ^{(*)}\to (f_1 \bar{f}_1) (f_2 \bar{f}_2)$ with a 
general CP-violating coupling {\it c.f.} Eq.~(\ref{param}),
we present here the differential distribution in the angle $\phi$ between
the planes of the two fermion pairs coming from the decays of the $Z^{(*)}$ 
bosons, taking into account the full dependence on the form factors $a,b$
and $c$. The notation is as fixed in the text.
\br
\frac{d\Gamma}{d\phi} \sim b_1 + b_2 \cos\phi + b_3 \sin\phi +
b_4 \cos 2\phi + b_5\sin 2\phi \;,  
\er
where
\br
b_1 &=& a^2 (2+\gamma_a^2) + |b|^2 x^2 \gamma_b^4 + 
8 \, |c|^2 \,x^2 \gamma_b^2 + 2 a\Re e(b) \, x \gamma_a \gamma_b^2
\nonumber\\
b_2 &=& -\frac{9\pi^2}{32} \eta_1\eta_2 \left[
a^2 \gamma_a
+ a\Re e(b) \,x \gamma_b^2 \right]
\nonumber\\
b_3 &=& \frac{9\pi^2}{16} \eta_1\eta_2 \left[
\Re e(b^*c) \, x^2 \gamma_b^3 
+ a\Re e(c) \, x \gamma_a\gamma_b \right]
\nonumber\\
b_4 &=& \frac{a^2}{2} - 2 \,|c|^2\, x^2 \gamma_b^2 \nonumber\\
b_5 &=& -2a\Re e(c) \,x \gamma_b \;.
\er
The polar angular distribution in $\theta_1$ is given by
\br
\frac{d\Gamma}{d\cos\theta_1} &\sim& a^2 [(\gamma_a^2-1) \sin^2 \theta_1 +2]
+ |b|^2 x^2 \gamma_b^4 \sin^2\theta_1 + 4|c|^2 x^2  \gamma_b^2 
(1+\cos^2\theta_1) \nonumber\\
&+& 2a\Re e(b) x \gamma_a \gamma_b^2 \sin^2\theta_1 - 8 a\Im m(c) \eta_1 x
\gamma_b \cos\theta_1\;.
\er



\end{document}